\documentclass[a4paper,11pt]{article} 
\pdfoutput=1

\usepackage{graphicx,rotating,hyperref,slashed,amsmath,xcolor,amssymb,amsfonts,colortbl,cite,cleveref,booktabs,soul} 
\makeatletter   
\hypersetup{colorlinks,bookmarksopen,bookmarksnumbered,
linkcolor=blus,pdfstartview=FitH,urlcolor=rossos,citecolor=verde}
\allowdisplaybreaks	 
\numberwithin{equation}{section} 
  
\hypersetup{%
    ,urlcolor=black
    ,citecolor=black
    ,linkcolor=black 
    }
\usepackage{mciteplus}

\AtBeginDocument{
  \hypersetup{
    urlcolor=blue,
    citecolor=blue,
    linkcolor=blue,
  }%
}

\newcommand{\hv}[1]{\ensuremath{\hat{#1}}}
\newcommand{\hn}{\ensuremath{\hat{n}}}
\newcommand{\hp}{\ensuremath{\hat{p}}}
\newcommand{\hq}{\ensuremath{\hat{q}}}

\newcommand{\qev}[1]{\ensuremath{\langle {#1} \rangle}}
\newcommand{\be}{\begin{equation}}
	\newcommand{\ee}{\end{equation}}
\newcommand{\bea}{\begin{eqnarray}}
	\newcommand{\eea}{\end{eqnarray}}
\newcommand{\barr}{\begin{array}}
	\newcommand{\earr}{\end{array}}
\newcommand{\ba}{\begin{align}}
	\newcommand{\ea}{\end{align}}

\newcommand{\lc}{\ensuremath{\mathcal{L}}}

\newcommand{\rc}{\ensuremath{\mathcal{R}}}
\newcommand{\fc}{\ensuremath{\mathcal{F}}}

\newcommand{\SNR}{\text{SNR}}

\def\bea{\begin{eqnarray}}
\def\eea{\end{eqnarray}} 

\def\be{\begin{equation}}
\def\ee{\end{equation}} 
\def\beq{\begin{equation}}
\def\eeq{\end{equation}}

\newcommand\ees{\end{eqnarray}}
\newcommand\bees{\begin{eqnarray}}

\def\bea{\begin{eqnarray}}
\def\eea{\end{eqnarray}}

\def\0{{\boldsymbol 0}}

\def\lsim{\mathrel{\rlap{\lower3pt\hbox{\hskip0pt$\sim$}}
   \raise1pt\hbox{$<$}}}         
\def\gsim{\mathrel{\rlap{\lower4pt\hbox{\hskip1pt$\sim$}}
   \raise1pt\hbox{$>$}}}         

 \newcommand{\sfootnote}[1]{}
\definecolor{bluc}{cmyk}{1,1,0,0.1}
\definecolor{rossoCP3}{cmyk}{0,.88,.77,.40}
\definecolor{rosso}{cmyk}{0,1,1,0.4}
\definecolor{rossos}{cmyk}{0,1,1,0.55}
\definecolor{rossoc}{cmyk}{0,1,1,0.2}
\definecolor{verdes}{cmyk}{0.92,0,0.59,0.4}

\hypersetup{colorlinks, bookmarksopen, bookmarksnumbered,
citecolor=verdes, linkcolor=bluc, pdfstartview=FitH, urlcolor=rossos}

\usepackage{multicol}
\usepackage{color}
\definecolor{rosso}{cmyk}{0,1,1,0.4}
\definecolor{rossos}{cmyk}{0,1,1,0.55}
\definecolor{rossoc}{cmyk}{0,1,1,0.2}
\definecolor{blu}{cmyk}{1,1,0,0.3}
\definecolor{blus}{cmyk}{1,1,0,0.6}
\definecolor{bluc}{cmyk}{1,1,0,0.1}
\definecolor{verde}{cmyk}{0.92,0,0.59,0.25}
\definecolor{verdec}{cmyk}{0.92,0,0.59,0.15}
\definecolor{verdes}{cmyk}{0.92,0,0.59,0.4}

\skip\@mpfootins = \skip\footins
\def\circa#1{\,\raise.3ex\hbox{$#1$\kern-.75em\lower1ex\hbox{$\sim$}}\,}

\newfam\rsfsfam
\def\mathscr#1{{\fam\rsfsfam\relax#1}}

\def\ev{\epsilon_V}

\def\circa#1{\,\raise.3ex\hbox{$#1$\kern-.75em\lower1ex\hbox{$\sim$}}\,}
\makeatletter

\def\hhref#1{\href{http://arxiv.org/abs/#1}{arXiv:#1}} 

\newcommand{\doi}[1]{\href{http://dx.doi.org/#1}{[doi]}}

\def\hhref#1{\href{http://arxiv.org/abs/#1}{arXiv:#1}} 
 
\def\art{\@ifnextchar[{\eart}{\oart}}
\def\eart[#1]#2#3#4#5#6{{\rm #2}, {\em #3 \bf #4} {\rm (#6) #5} ({\em #1})}

\def\article{\@ifnextchar[{\earticle}{\oarticle}}
\def\oarticle#1#2#3#4#5#6{{\rm #1}, {\em ``#6''}, {\rm #2 #3 (#5) #4}}
\def\earticle[#1]#2#3#4#5#6#7{{\rm #2}, {\em ``#7''}, {\rm #3 #4 (#6) #5}  [\hhref{#1}]}
\def\hepart[#1]#2{{\rm #2, \em#1}}
\def\heparticle[#1]#2#3{#2, {\em ``#3''} [\hhref{#1}]}

\newcounter{alphaequation}[equation]
\def\thealphaequation{\theequation\hbox to
0.6em{\hfil\alph{alphaequation}\hfil}}
\def\eqnsystem#1{
\def\@eqnnum{{\rm (\thealphaequation)}}
\def\@@eqncr{\let\@tempa\relax \ifcase\@eqcnt \def\@tempa{& & &} \or
  \def\@tempa{& &}\or \def\@tempa{&}\fi\@tempa
  \if@eqnsw\@eqnnum\refstepcounter{alphaequation}\fi
\global\@eqnswtrue\global\@eqcnt=0\cr}
\refstepcounter{equation} \let\@currentlabel\theequation \def\@tempb{#1}
\ifx\@tempb\empty\else\label{#1}\fi
\refstepcounter{alphaequation}
\let\@currentlabel\thealphaequation
\global\@eqnswtrue\global\@eqcnt=0 \tabskip\@centering\let\\=\@eqncr
$$\halign to \displaywidth\bgroup \@eqnsel\hskip\@centering
$\displaystyle\tabskip\z@{##}$&\global\@eqcnt\@ne
\hskip2\arraycolsep\hfil${##}$\hfil& \global\@eqcnt\tw@\hskip2\arraycolsep
$\displaystyle\tabskip\z@{##}$\hfil
\tabskip\@centering&\llap{##}\tabskip\z@\cr}
\def\endeqnsystem{\@@eqncr\egroup$$\global\@ignoretrue} \makeatother

\definecolor{fiorentina}{rgb}{.5,0,.5}

\begin{document}

\setcounter{page}{1} \baselineskip=15.5pt \thispagestyle{empty}

\vspace{0.8cm}
\begin{center}

{\fontsize{19}{28}\selectfont  \sffamily \bfseries {Measuring  the circular polarization \\ of gravitational waves with 
pulsar timing arrays
}}

\vspace{0.2cm}

\begin{center}
{\fontsize{12}{30}\selectfont  
N.~M.~Jim\'enez Cruz$^{a}$ \footnote{\texttt{nmjc1209.at.gmail.com}}, Ameek Malhotra$^{a}$ \footnote{\texttt{ameek.malhotra.at.swansea.ac.uk}}, Gianmassimo Tasinato$^{a, b}$ \footnote{\texttt{g.tasinato2208.at.gmail.com}}, Ivonne Zavala$^{a}$ \footnote{\texttt{e.i.zavalacarrasco.at.swansea.ac.uk}}
} 
\end{center}

\begin{center}

\vskip 8pt
\textsl{$^{a}$ Physics Department, Swansea University, SA28PP, United Kingdom}\\
\textsl{$^{b}$ Dipartimento di Fisica e Astronomia, Universit\`a di Bologna,\\
 INFN, Sezione di Bologna, I.S. FLAG, viale B. Pichat 6/2, 40127 Bologna,   Italy}
\vskip 7pt

\end{center}

\smallskip
\begin{abstract}
\noindent
{The circular polarization of the stochastic gravitational wave background (SGWB) is a key observable for characterising the origin of the  signal detected by Pulsar Timing Array (PTA) collaborations. Both the astrophysical and the cosmological SGWB can have a sizeable amount of circular polarization, due to Poisson fluctuations in the source properties for the former, and  to parity violating processes in the early universe for the latter. Its measurement is  challenging since  PTA are blind to the circular polarization monopole, forcing us to turn to  anisotropies for detection.  We study the sensitivity of current and future PTA datasets to circular polarization anisotropies, focusing on realistic modelling of  intrinsic and  kinematic anisotropies for  astrophysical and cosmological scenarios respectively. Our results indicate that the expected level of circular polarization for the astrophysical SGWB should   be within the reach of near future datasets, while for cosmological SGWB circular polarization is a viable target for more advanced SKA-type experiments.}
\end{abstract}

\end{center}

\section{Introduction}

The recent  hints of detection of a stochastic gravitational
wave background (SGWB) by several pulsar timing arrays (PTA) collaborations \cite{NANOGrav:2023gor,Reardon:2023gzh,Xu:2023wog,EPTA:2023fyk,InternationalPulsarTimingArray:2023mzf} 
raise questions about its origin,  whether astrophysical or cosmological: see respectively
\cite{Sesana:2008mz,Burke-Spolaor:2018bvk} and \cite{Caprini:2018mtu} for reviews, and 
\cite{Babak:2024yhu} for a recent assessment on the challenges
to distinguish among different SGWB sources with PTA experiments. 
The amount of circular polarization imprinted
in the SGWB constitutes 
a possible discriminator
among the two possibilities. Several early universe sources can induce a circularly
polarized SGWB, for example associated with phenomena of parity violation in gravitational
or vector sectors, see e.g.  \cite{Jackiw:2003pm,Lue:1998mq,Satoh:2007gn,Contaldi:2008yz,Alexander:2009tp,Anber:2012du,Bartolo:2016ami,Dimastrogiovanni:2016fuu,Nishizawa:2018srh}.
Moreover, an astrophysical   SGWB  is also characterized by  
a sizeable amount of circular polarization, related with  the distribution of inclination angles of black hole
binary sources \cite{ValbusaDallArmi:2023ydl,Sato-Polito:2023spo}. 
But a measurement  of  circular 
polarization with  PTA, although interesting, is  challenging  since the isotropic part of the background is {\it insensitive} to this
quantity \cite{Kato:2015bye}. The reason being the geometrical configuration of the PTA system, which makes its response to circular polarization vanish. This feature is in common with what happens
for planar GW interferometric detectors, such as LISA (see e.g. \cite{Seto:2006hf,Smith:2016jqs,Domcke:2019zls,LISACosmologyWorkingGroup:2022jok,Colpi:2024xhw}).
Hence, an option  for detecting the SGWB circular 
polarization with PTA is to measure  SGWB anisotropies: see e.g. 
\cite{Kato:2015bye,Hotinli:2019tpc,Belgacem:2020nda,Sato-Polito:2021efu,Tasinato:2023zcg}, and \cite{Romano:2016dpx} for a review. 
For
cosmological sources, anisotropies
are induced by early universe effects, and an approach  to describe  SGWB anisotropies and their properties in this context  has been recently developed (see e.g. \cite{Alba:2015cms,Contaldi:2016koz,Geller:2018mwu,Bartolo:2019oiq,Bartolo:2019zvb,Bartolo:2019yeu,Adshead:2020bji,DallArmi:2020dar,
LISACosmologyWorkingGroup:2022kbp,Malhotra:2022ply,Schulze:2023ich}).  For astrophysical
sources,  SGWB anisotropies arise due to clustering of galaxies where the GW sources reside, as well as Poisson-type fluctuations~\cite{Cornish:2013aba,Mingarelli:2013dsa,Taylor:2013esa,Mingarelli:2017PTA,Cornish:2015ikx,Taylor:2020zpk,Becsy:2022pnr,Allen:2022dzg,Sato-Polito:2021efu,Sato-Polito:2023spo,NANOGrav:2023tcn,Sah:2024oyg}. 
 SGWB anisotropies  have 
not been measured yet with PTA experiments \cite{Taylor:2015udp,NANOGrav:2023tcn}, but
given their rich physics content 
their detectability represents  an active an interesting line of investigation  -- see e.g. \cite{Anholm:2008wy,Taylor:2013esa,Mingarelli:2013dsa,Gair:2014rwa,Cornish:2014rva,Mingarelli:2017fbe,Hotinli:2019tpc,Ali-Haimoud:2020ozu,Ali-Haimoud:2020iyz,Pol:2022sjn,Bernardo:2023jhs,Nay:2023pwu}.

The aim of this work is to forecast
the sensitivity of PTA experiments
to circular polarization, making use
for the first time 
of realistic modelling of SGWB anisotropies.
We study the two options of cosmological and astrophysical SGWB.

For the first  case of GW sources from the early universe, the largest contribution
to  SGWB anisotropies is associated to kinematic Doppler effects, 
due to our motion with respect to the source of SGWB~\cite{LISACosmologyWorkingGroup:2022kbp,Cusin:2022cbb}. 
They are the GW analog to kinematic effects measured in the cosmic
microwave background radiation~\cite{Smoot:1977bs,Kogut:1993ag,WMAP:2003ivt,WMAP:2008ydk,Planck:2013kqc}. Kinematic anisotropies in the SGWB are  deterministically  controlled by   the isotropic part of the background, and the amplitude of the kinematic dipole  
is expected to be of order ${\cal O}(10^{-3})$ smaller than the isotropic monopole. The PTA response to kinematic anisotropies  can be analytically
computed~\cite{Anholm:2008wy,Mingarelli:2013dsa,Tasinato:2023zcg}, including the effects of circular polarization, and  modified gravity signatures~\cite{Tasinato:2023zcg}. In \cite{Cruz:2024svc} we forecasted the sensitivity of future PTA experiments to kinematic effects related to the SGWB intensity. 
In the present work, sections \ref{sec_theory}, \ref{sec_present} and \ref{sec_fut},
we study using different approaches how the measurement of kinematic anisotropies can be used to detect the circular polarization of the cosmological
background with future PTA data. We find that monitoring a large number
of monitored pulsars will be necessary to detect the signal
we are interested in.

A more optimistic result is obtained
investigating
the  case of astrophysical SGWB, studied in
 section \ref{sec_astro}. For such a GW signal,
%
the degree
of intrinsic anisotropy is sizeable, of order up to \mbox{${\cal O}(10^{-1}\text{--}10^{-2})$} relative to the monopole, hence much larger than the kinematic
one~\cite{Cornish:2013aba,Mingarelli:2013dsa,Taylor:2013esa,Mingarelli:2017PTA,Cornish:2015ikx,Taylor:2020zpk,Becsy:2022pnr,Allen:2022dzg,Sato-Polito:2021efu,
Sato-Polito:2023spo}. Interestingly, the astrophysical SGWB is expected
to be strongly circularly polarized \cite{ValbusaDallArmi:2023ydl,Sato-Polito:2023spo}, with an amplitude of circular polarization comparable to the SGWB intensity -- this property of course improves the prospects of detection. In fact,  our analysis suggests that
a monitoring few hundred pulsars (which should be achievable
in the SKA era \cite{Janssen:2014dka,Keane:2014vja,Weltman:2018zrl}) should be sufficient for detecting
circular polarization in the astrophysical SGWB. 

We conclude in section \ref{sec_concl}.  We adopt units in which 
 $c=\hbar=1$.

\section{ Our set-up}
\label{sec_theory}

We start with a theoretical section, based on \cite{Tasinato:2023zcg},  reviewing the  formulas needed
for our analysis. We focus  in this section on kinematic
anisotropies only, with the amplitude of the kinematic dipole of order ${\cal O}(10^{-3})$ with respect to the background monopole. We then  have in mind 
cosmological sources for the SGWB,  as explained in the introduction.
Other works discussing kinematic effects on SGWB include \cite{LISACosmologyWorkingGroup:2022kbp,Cusin:2022cbb,Bertacca:2019fnt,ValbusaDallArmi:2022htu,Chung:2022xhv,Chowdhury:2022pnv,Cusin:2024git}. A discussion
of intrinsic anisotropies for astrophysical backgrounds, and the associated measurement of circular polarization, can be found in section \ref{sec_astro}.

Gravitational
waves (GW) are fluctuations $h_{ij}(t, \vec x)$ around
the Minkowski metric 
\be
d s^2\,=\,-d t^2+\left[ \delta_{ij}+h_{ij}(t, \vec x)
\right]\,d x^i d x^j
\,.
\ee
They are
decomposed in Fourier modes as 
\be
\label{fouhij}
h_{ij}(t,\vec x)\,=\,\sum_{\lambda}\,\int_{-\infty}^{+\infty} d f\,\int d^2 \hat n\,{ e}^{-2 \pi i f \,\hat n \vec x}\,e^{2 \pi i f t}\,
{\bf e}_{ij}^\lambda (\hat n)\,h_{\lambda}(f, \hat n)\,,
\ee
 imposing the condition 
$h_\lambda (-f, \hat n)\,=\,h^*_\lambda (f, \hat n)
$, 
  which ensures that $h_{ij}(t, \vec x)$ is real.
We adopt a $\lambda=(+,\times)$ basis
for the polarization tensors
 ${\bf e}_{ij}^{\lambda}$  
   in  eq \eqref{fouhij};
   we assume they are real.

Pulsar timing arrays (PTA) measure GW through the time delay they cause on the observed pulsar period. Such
time delays are associated
with GW deformations  of light geodesics from emission
to detection. We denote pulsars with latin
letters: $a$, $b$, etc. The time delay associated with 
pulsar $a$ is denoted with  $z_a(t)\,=\,{\Delta T_a(t)}/{T_a(t)}$. We decompose it in Fourier modes as  
\bea
z_a(t)&=&
 \int_{-\infty}^{+\infty} d f\,e^{2 \pi i f t}\,z_a(f)
 \nonumber
\\
&=&
\int_{-\infty}^{+\infty} d f\,e^{2 \pi i f t}\,\left(\sum_\lambda\,\int d^2 \hat n\,
D_a^{ij}(\hat n)\,{\bf e}_{ij}^\lambda (\hat n)\,h_{\lambda}(f, \hat n)
\right)\,.
\label{deftd}
\eea
\bigskip

\noindent
The  detector tensor $D_a^{ij}(\hat n)$ is defined 
 in terms of polarization tensors
as
$
2 {(1+\hat n \cdot \hat x_a)}\, D^{ij}_a\equiv {\hat x_a^i\,\hat x_a^j}
$. The quantity $\vec x_a\,=\,\tau_a\,\hat x_a$ is the pulsar position, and 
$\tau_a$ is the light travelling time
from emission to detection.

We express the GW two-point correlators as
(we use notation of \cite{Smith:2016jqs,Smith:2019wny}):
\be
\label{corrh1}
\langle
h_{\lambda}(f, \hat n)\,h^*_{\lambda'}(f', \hat n')
 \rangle 
 \,=\,\frac12\,S_{\lambda \lambda'}(f,\hat n)\,\delta(f-f')\,\frac{\delta^{(2)} (\hat n-\hat n')}{4\pi}
\,.
\ee
 The quantity  $S_{\lambda \lambda'}(f,\hat n)$  is decomposed into intensity and circular polarization: 
\be
\label{decsa}
S_{\lambda \lambda'}(f,\hat n)\,=\,
I(f,\hat n) \delta_{\lambda \lambda'}-i V(f, \hat n)\,\epsilon_{\lambda \lambda'}
\,,
\ee
where the $2\times2$ tensor $\epsilon_{\lambda \lambda'}$ components are  $\epsilon_{+ \times}\,=\,1\,=\,-\epsilon_{\times+}$, while $\epsilon_{+ +}\,=\,0\,=\,\epsilon_{\times \times}$. 

The SGWB intensity $I(f,\hat n) $ is real and positive.  The circular polarization $V(f,\hat n) $ is a real quantity. 
Both quantities can depend on the GW frequency and direction, and behave as scalars under boosts. 
We introduce the short-hand  notation 
\be
D_a^{\lambda}(\hat n)\,\equiv\,
D_a^{ij}(\hat n)\,\,{\bf e}_{ij}^\lambda (\hat n)\,,
\hskip1cm;\hskip1cm
\Delta t_{12}\,=\,t_1-t_2\,,
\ee
and  compute the two-point correlator between  two pulsar 
time-delays, associated with  a pulsar pair $a$ and $b$. We 
find  (we sum over repeated indexes)
\bea
\langle z_a(t_1) z_b(t_2)
\rangle
&=&\frac12 \int{ d f}  \,d^2 \hat n
\,D_a^{\lambda}(\hat n) D_b^{\lambda'}(\hat n)
\left[ \cos{(2 \pi  f \Delta t_{12})}\, I (f,\hat n) \delta_{\lambda \lambda'} 
+\sin{(2 \pi  f \Delta t_{12})}\, V(f,\hat n) \epsilon_{\lambda \lambda'} 
\right]\,.
\nonumber\\
\eea
This formula indicates that  circular polarization is only detectable at unequal time measurements: $\Delta t_{12}\neq0$
(see e.g. \cite{Domcke:2019zls}). 
It is convenient  to make use of
  time residuals
\be
\label{deftra}
R_a(t)\,\equiv\,\int_0^t\,d t' z_a(t')\,,
\ee
a quantity  easier to handle when Fourier transforming the signal. Their
two-point function
is
\bea
&&
\langle 
R_a(t_A)
R_b(t_B)
\rangle
=\frac12 \int_0^{t_A} \int_0^{t_B}\,d t_1 d t_2\,\int d f \,d^2 \hat n
\nonumber
\\
&&
\times
\Big[ \cos{\left(2 \pi  f \Delta t_{12} \right)}\, 
\gamma_{ab}^{I}(f,\hat n)\,
I (f,\hat n) \delta_{\lambda \lambda'} 
+\sin{\left(2 \pi  f \Delta t_{12} \right)}\, 
\gamma_{ab}^{I}(f,\hat n)\,
V(f,\hat n) 
\Big]
\label{twopcordtA}
\\
&=&
\int \frac{d f\, \sin{\left( \pi f t_A \right)}\sin{\left( \pi f t_B \right)}}{\pi\,f^2}
\left[\bar I(f)\,
\Gamma^I_{ab}(f)\,\cos{\left(2 \pi  f \Delta t_{AB} \right)}\,
+\bar V(f)\,\Gamma^V_{ab}(f)\,\sin{\left(2 \pi  f \Delta t_{AB} \right)}
\right]\,,
\label{twopcordtaB}
\nonumber
\\
\eea
where $\bar I(f)$  is the isotropic value of the intensity integrated over all directions, while $\bar V(f)$
is its analog for circular polarization.
In writing eq \eqref{twopcordtA} we introduce 
\bea
\gamma_{ab}^{I}&=&
D_a^{\lambda}(\hat n) D_b^{\lambda'}(\hat n)\,\delta_{\lambda \lambda'}
\nonumber
\\
&=&
\frac{(\hat x_a \cdot \hat n)^2+(\hat x_b \cdot \hat n)^2+ (\hat x_a \cdot \hat n)^2  (\hat x_b \cdot \hat n)^2-1}{8(1+\hat x_a \cdot \hat n)(1+\hat x_b \cdot \hat n)}
\nonumber\\
&+&\frac{ (\hat x_a \cdot \hat x_b)^2-2 (\hat x_a \cdot \hat x_b)  (\hat x_a \cdot \hat n)  (\hat x_b \cdot \hat n) 
}{4(1+\hat x_a \cdot \hat n)(1+\hat x_b \cdot \hat n)}
\,,
\eea
and 
\bea
\label{def_gav}
\gamma_{ab}^{V}&=&
D_a^{\lambda}(\hat n) D_b^{\lambda'}(\hat n)\,\epsilon_{\lambda \lambda'}
\nonumber
\\
&=&
\frac{\left[ \hat x_a \cdot \hat x_b 
- (\hat x_a \cdot \hat n)  (\hat x_b \cdot \hat n) 
\right] \left[ \hat n \cdot ( \hat x_a \times \hat x_b ) \right] }{{4(1+\hat x_a \cdot \hat n)(1+\hat x_b \cdot \hat n)}}
\,,
\eea
with $\cdot$ denoting dot product, and $\times$  cross product among vectors.   
In passing
from eq \eqref{twopcordtA} to eq \eqref{twopcordtaB} we integrate along directions, and
we define the PTA response functions
\bea
\label{gammai1}
\Gamma^I_{ab}(f)&=&\frac{1 }{{2 \pi\,\bar I(f)}}\int d^2 \hat n\left(
D_a^{\lambda}(\hat n) D_b^{\lambda'}(\hat n)\,\delta_{\lambda \lambda'} \right)\,
{I (f,\hat n)} 
\,,
\\
\label{gammav1}
\Gamma^V_{ab}(f)&=&\frac{1 }{2\pi\,{\bar V(f)}}\int d^2 \hat n\left(
D_a^{\lambda}(\hat n) D_b^{\lambda'}(\hat n)\,\epsilon_{\lambda \lambda'} \right)\,
{V (f,\hat n)} 
\,.
\eea
The response functions are key quantities to study the response
of the PTA system to GW observables.

We can avoid to carry on the oscillating functions in the integrand of eq \eqref{twopcordtaB},
and define
\bea
\langle 
R_a
R_b
\rangle
&\equiv&
\int \frac{d f}{\pi\,f^2}
\,
\left[\bar I(f)\,
\Gamma^I_{ab}(f)\,
+\bar V(f)\,\Gamma^V_{ab}(f)
\right]\,,
\label{twopcordt}
\eea
In fact, 
 the quantities $t_A$, $t_B$ in
 eq \eqref{twopcordtaB} indicate that 
 we are measuring pulsar signals at different times, separated
 by an interval corresponding to the
  `cadence' of detection. PTA
collaborations
 measure
pulsar periods around once per week: the cadence then establishes
the upper bound in frequency of PTA measurements $50$ yr$^{-1}\sim 10^{-6}$ Hz. 
  Hence, we  typically
  work in a regime where $\left(f t_A, \,f t_B, \,f  \Delta t_{AB} \right) \sim {\cal O}(1)$.  Consequently,  the oscillating functions in the
  integrand of eq \eqref{twopcordtaB} are expected to give order one factors, and
  will  be neglected in the definition of the two-point function
  passing
  from \eqref{twopcordtaB} to \eqref{twopcordt}.
   See e.g. \cite{Maggiore:2018sht}, chapter 23.

\smallskip

While so far our formulas are  general, we now focus on kinematic anisotropies,
following the treatment of \cite{Cruz:2024svc,Tasinato:2023zcg}. Our
motion with velocity $\vec v\,=\,\beta \hat v$ with respect to the SGWB rest frame,  induces
kinematic anisotropies in intensity $I$ and circular
polarization $V$. They are 
 expressed as 
\bea
\frac{I(f,\hat n)}{\bar I(f)}
&=&{\cal D}\,\frac{\bar I({\cal D}\,f )}{\bar I(f)}\,,
\hskip0.8cm
\hskip0.8cm
\frac{V(f,\hat n)}{\bar V(f)}\,=\,{\cal D}\,\frac{\bar V({\cal D}\,f )}{\bar V(f)}
\label{genekaa}\,,
\eea
with
\be
\label{geneka}
{\cal D}\,=\,\frac{\sqrt{1-\beta^2}}{1-\beta \hat n \cdot \hat v}\,,
\ee
where $\hat n$ is the GW direction, and $\hat v$
the relative velocity among frames. 
Since in this and the next section we are assuming a cosmological
origin for the SGWB, we expect $\beta$ to be of the 
same order of the value measured by cosmic
microwave background: $\beta\,=\,1.2 \times 10^{-3}$ \cite{Smoot:1977bs,Kogut:1993ag,WMAP:2003ivt,WMAP:2008ydk,Planck:2013kqc}. With $\beta$ being
 small, we can expand both quantities
in \eqref{genekaa} at first order in $\beta$, and plug
the resulting expressions in the response
functions of eqs \eqref{gammai1}, \eqref{gammav1}. 
 In this way, we determine the PTA response to the kinematic
 dipole in intensity and in circular polarization. 
Defining $n_I\,=\,{d\,\ln \bar I(f)}/{d\,\ln f}$, $n_V\,=\,{d\,\ln \bar V(f)}/{d\,\ln f}$, we
 obtain

\begin{figure}[t!]
	\centering
	\includegraphics[width=0.45\linewidth]{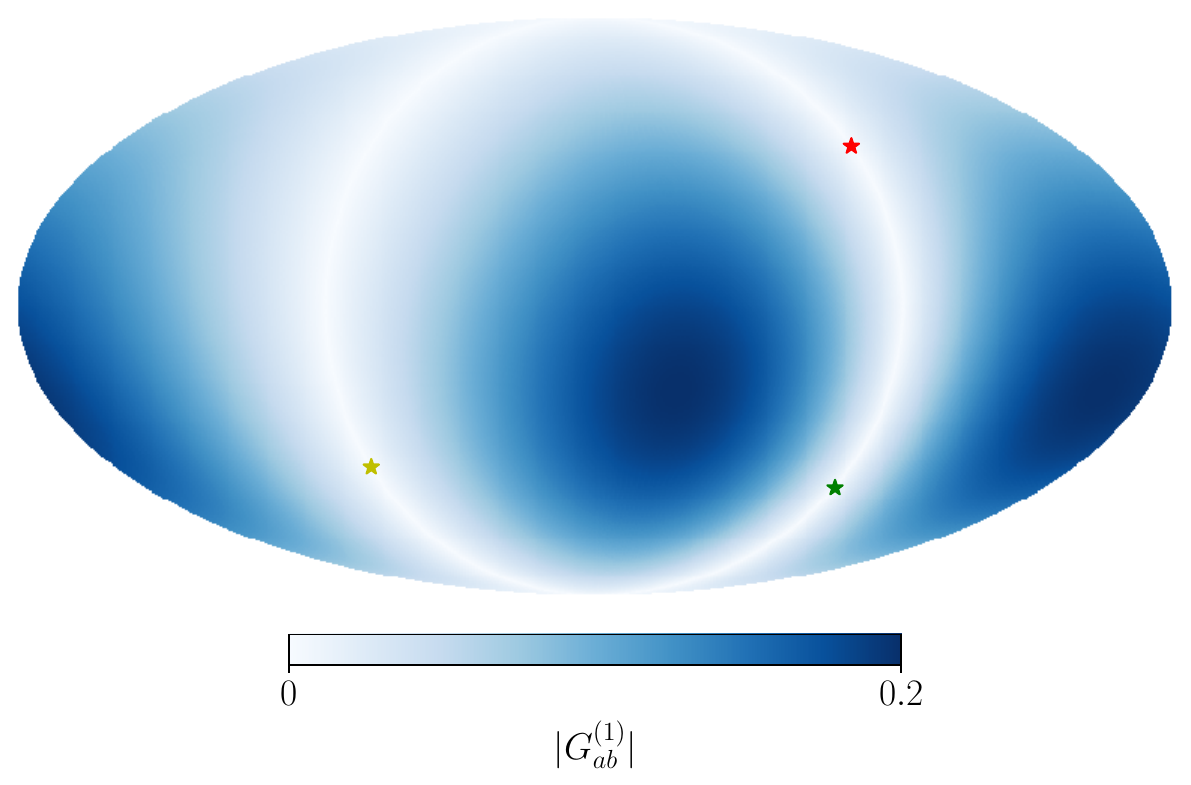}
 	\includegraphics[width=0.45\linewidth]{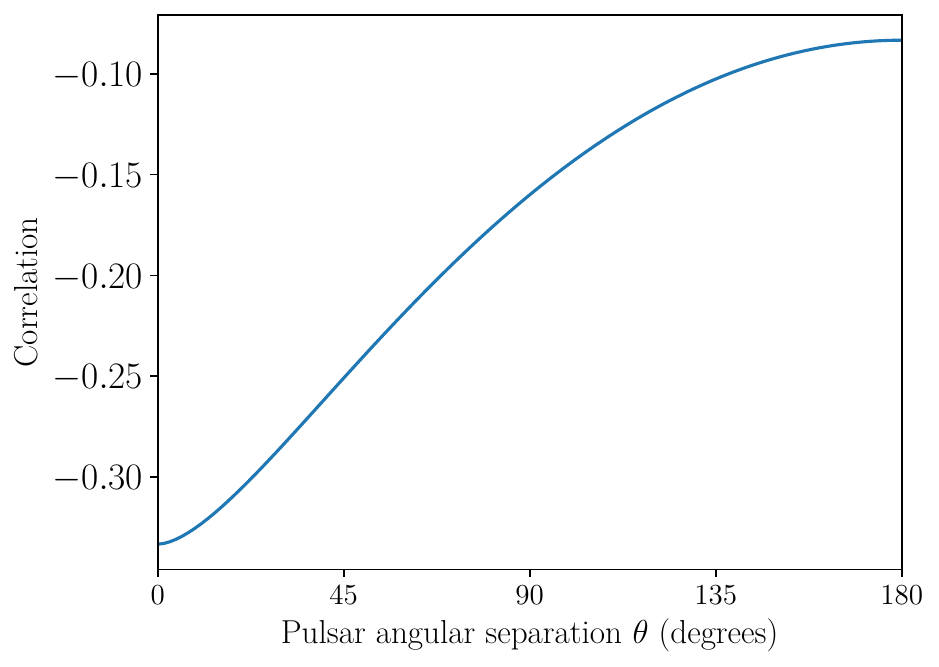}
	\caption{  \it{{\bf Left:} the magnitude of the PTA response function  $G^{(1)}_{ab}$  to kinematic dipole anisotropies in circular polarization, see eq \eqref{defgv1}. We fix the velocity vector $\hat v$ along the the direction measured by the CMB ($\hat v$ and $-\hat v$ are denoted by red and yellow stars respectively). We plot the response as a function of the positions
of a pair of pulsars, with one pulsar fixed to a direction perpendicular to $\hat v$ (green star). %
   {\bf Right:} the dipole response as a function of the angle between the pulsars, without  including the $\hat v\cdot (\hat x_a\times \hat x_b)$ factor in eq  \eqref{defgv1}.}}
	\label{fig:response_V}
\end{figure}
\bea
\label{GammaABI}
\Gamma_{ab}^{I}&=&\Gamma_{ab}^{(0)}+\beta\,\left(n_I-1\right)\,\Gamma_{ab}^{(1)}\,,
\\
\label{GammaABV}
\Gamma_{ab}^{V}&=&\beta\,\left(n_V-1\right)\,G_{ab}^{(1)}\,,
\eea
with
\bea
\Gamma_{ab}^{(0)}&=&\frac13-\frac{y_{ab}}{6}+y_{ab} \ln y_{ab} \,,
\\
\Gamma_{ab}^{(1)}&=&\left(\frac{1}{12}+\frac{ y_{ab}}{2}+\frac{ y_{ab} \ln y_{ab}}{2(1-y_{ab})} \right)
\, \left[\hat v\cdot \hat x_a+\hat v\cdot \hat x_b\right]\,,
\label{defgi1}
\\
G_{ab}^{(1)}&=&-\left(\frac13+\frac{y_{ab} \ln y_{ab}}{4(1-y_{ab})} \right)\,\left[ \hat v\cdot (\hat x_a\times \hat x_b)\right]\,,
\label{defgv1}
\eea
with 
$
y_{ab}\,=\,\left(1-\hat x_a  \cdot \hat x_b
\right)/2
$. 
As stated in  the introduction, the PTA
response to circular polarization $\Gamma_{ab}^{V}$ vanishes for 
an isotropic background.  In fact, the response of eq \eqref{GammaABV}  is non-zero only in the presence of anisotropies, since it starts at first order in the $\beta$ expansion.

While $\Gamma_{ab}^{(0)}$ corresponds to the well-known Hellings-Downs response to the isotropic part of the background, the
 two contributions $\Gamma_{ab}^{(1)}$
 and $G_{ab}^{(1)}$ control the PTA response
 to the kinematic dipole (respectively to the intensity
 $I$ and circular polarization $V$ of the GW). 

 Interestingly, while the PTA
 response is more sensitive to the intensity kinematic dipole when pulsar vectors lie {\it parallely} to the 
 velocity vector $\hat v$ (see eq \eqref{defgi1} --
 we studied this phenomenon in \cite{Cruz:2024svc}), the
 opposite is true for the circular polarization
 dipole \cite{Tasinato:2023zcg}, which can be measured more easily with pulsars
located 
{\it  orthogonally} to $\hat v$ (see eq \eqref{defgv1}). See e.g. Fig \ref{fig:response_V}
for a graphical representation of the quantity $G^{(1)}_{ab}$. 
This property implies that an appropriate choice of the pulsars to measure -- depending on their position -- can make enhance
the sensitivity on the signal. 
We will make
use of this feature
in what follows.

These results complete the review of  the theoretical
tools we can apply to the forecast of detectability
of circular polarization with PTA experiments,
a topic we  discuss next.

\section{Circular polarization and present-day PTA experiments}
\label{sec_present}

After reviewing and developing 
in section \ref{sec_theory} 
the
theoretical basis for our work, we  discuss practical
strategies for detecting
circular polarization with PTA data. As we learned,
the detection of circular polarization
is tied with the measurement of SGWB anisotropies.

We  forecast the prospects to detect
the SGWB
circular polarization $V$ by measuring the kinematic dipole
of the SGWB with PTA. In this section
we have in mind a 
cosmological SGWB which can  have a potentially  large amplitude of circular
polarization, as large
as the SGWB intensity, see e.g. \cite{Jackiw:2003pm,Lue:1998mq,Satoh:2007gn,Contaldi:2008yz,Alexander:2009tp,Anber:2012du,Bartolo:2016ami,Nishizawa:2018srh,Figueroa:2023zhu,Geller:2023shn,Ellis:2023oxs,Unal:2023srk,Dimastrogiovanni:2023oid}.  
Interestingly, also the  astrophysical SGWB
is expected to be  characterized by a large 
amount of circular polarization: we discuss it
in section \ref{sec_astro}.

We proceed following two approaches. First, in this
section, we focus on  near future PTA experiments
in the SKA era \cite{Janssen:2014dka,Keane:2014vja,Weltman:2018zrl},
which will monitor a finite number (of order few
hundreds) pulsars. 
Building on the results
of section  \ref{sec_theory}, we show how the sensitivity
to circular polarization can improve by choosing carefully  the 
 position in the sky of the monitored pulsars. It is very hard though to reach a level of 
 sensitivity compatible with a detection.
   In the
 next section \ref{sec_fut}, then, we  make some simplifying 
 assumptions to study the problem, and we examine the case of futuristic PTA experiments monitoring
 very large (of order thousands) pulsars. In such a case, the experiments will have enough sensitivity to detect the effects we are interested about.

 Other works discussing  prospects  of
detection of circular polarization with PTA, also following alternative methods,  are \cite{Kato:2015bye,Hotinli:2019tpc,Belgacem:2020nda,Sato-Polito:2021efu}. As far as we are aware, 
ours is the first work which makes use
of   realistic modelling of  SGWB anisotropies
 (both for cosmological and astrophysical
sources) towards the detection of $V$.

\smallskip

We assume
that the isotropic part of intensity $\bar I(f)$
and  circular polarization $\bar V(f)$ -- obtained
integrating over all angular directions -- follows a
power-law profile 
given by 
\bea
\label{intraiav}
\bar I(f)\,=\,
I_0
\left( \frac{f}{f_\star} \right)^{n_I}
\hskip0.7cm,\hskip0.7cm
\bar V(f)\,=\,V_0
\left( \frac{f}{f_\star} \right)^{n_V}
\eea
with $f_\star$ a reference frequency (we
take $f_\star\, =\,1$ yr$^{-1}$, as appropriate
for PTA experiments), while $n_I$
and $n_V$ are the  tilts defined above eq \eqref{GammaABI}. 
The constant quantities $ I_0$
and $ V_0$ control the amplitude
of intensity and circular polarization.
Following  the conventions of  \cite{NANOGrav:2023gor}
we 
introduce the constants $A_I$ and $A_V$
as
\be
\label{defofA}
 I_0\,=\,\frac{A_I^2}{2 f_\star} \hskip0.5cm,\hskip0.5cm
 V_0\,=\,\frac{A_V^2}{2 f_\star}
\ee
as alternative parameters to test. The definition \eqref{defofA}
allows us to compare more directly with the results of \cite{NANOGrav:2023gor}. 

\subsection{Limits from current datasets, and definition of the 
 likelihood}

We focus here on PTA experiments 
with a finite number of monitored pulsars. We take  the NANOGrav 
configuration  system \cite{NANOGrav:2023ctt,NANOGrav:2023hde,the_nanograv_collaboration_2023_8092346,the_nanograv_collaboration_2023_8423265} as reference. The NANOGrav 
collaboration collects
data   monitoring 67 pulsars.
First, we show that
the   sensitivity of current version 
of the NANOGrav pulsar system 
 is  {\it not sufficient} to  detect
a  signal associated with the SGWB circular polarization $V$.
To improve the situation, we then imagine 
  to
 add a finite number (up to few  hundred) of
 monitored pulsars  to the NANOGrav pulsar  set. For definiteness, we
 adopt the same noise curves of the NANOGrav pulsars
 \cite{NANOGrav:2023hde}, and we locate the additional objects
 at appropriate positions in the sky
  in order to be more
  sensitive to circular polarization. 

  To start with, 
 we  make use of publicly available tools and assess to what extent current can PTA data set useful constraints on  SGWB circular polarization. We analyse the NANOGrav data set~\cite{NANOGrav:2023gor}  (referred to as NG15 hereafter) looking for the presence of circular polarization through the kinematic dipole. We fix the dipole magnitude and direction to be the same as the cosmic
microwave background~\cite{WMAP:2008ydk}.    
We apply  the likelihood made available by the NANOGrav collaboration through the packages \texttt{ENTERPRISE}~\cite{enterprise} and \texttt{ENTERPRISE-EXTENSIONS} \cite{enterprise-ext}. We explore the parameter space using the \texttt{MCMC} sampler~\cite{Lewis:2002ah,Lewis:2013hha}, through the \texttt{Cobaya} interface~\cite{Torrado:2020dgo}. We vary the SGWB intensity amplitude $\log_{10}A_I$, the spectral tilt in the combination  
$$\gamma=2- n_I\,,$$
the relative circular polarization amplitude~\footnote{While the parameter $\ev$ runs
between -1 and 1, from now on in our Fisher forecasts we only will consider its absolute value $|\ev|$
and take it to be positive.} 
\be
\ev \equiv V_0/I_0 \in [-1,1]\,,
\ee
and we fix the individual pulsar red noise parameters to their median values from the NG15 analyses for computational ease.
The key parameter controlling the amplitude of circular polarization versus intensity  is $\epsilon_V$. 
\begin{figure}[ht]
    \centering
    \includegraphics[width=0.65\linewidth]{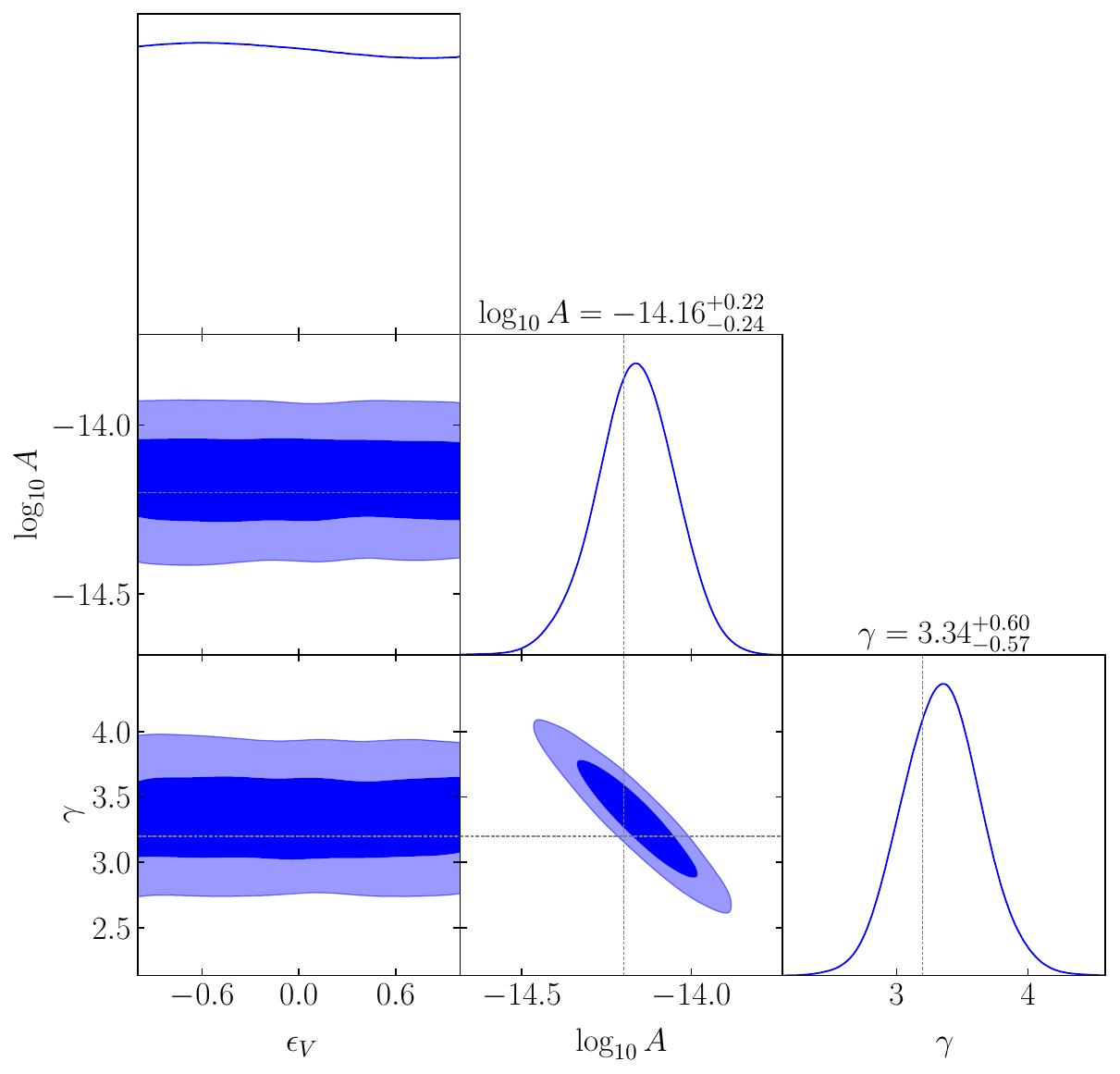}
    \caption{\textit{Marginalised 2D and 1D contours of the SGWB parameters obtained using the NG15 dataset. The flat distribution for $\ev$ indicates that current datasets are unable to put any constraints on the amplitude of circular polarization through the kinematic dipole, as explained in the main text. Dashed lines represent the median values from NG15 and the parameter limits shown correspond to $95\%$ limits.} }
    \label{fig:NG15_contours}
\end{figure}
We plot the marginalised distributions for the SGWB parameters in Fig
 \ref{fig:NG15_contours} using \texttt{GetDist}~\cite{Lewis:2019xzd}. The figure shows that current data  are 
in fact unable to put any constraints on the magnitude of circular polarization. The reason can be understood as follows. As we have learned above PTA systems are blind to the circular polarization monopole, thus we need
to measure the SGWB anisotropy to detect it. But, as shown in~\cite{Cruz:2024svc}, current data impose only  an upper limit $\beta < 0.3$  on the kinematic dipolar anisotropy. Thus, by fixing $\beta=1.23\times 10^{-3}$, $\epsilon_V$ can be freely varied within its prior range $[-1,1]$ without creating significant deviations from the observed inter-pulsar correlations.\footnote{Varying the individual pulsar red noise parameters will not change our results, since the circular polarization is essentially unconstrained at the moment.}

\smallskip

To improve on the current situation, a possibility is
to monitor additional
 pulsars appropriately located in 
the sky (following the criteria
of section \ref{sec_theory}), so to collect more information specifically
on the SGWB circular polarization. This
is the strategy we adopt in this section. To study
this option, 
we introduce  a Gaussian likelihood for intensity
and circular polarization, ${I}$ and ${V}$, making use of ideas first developed in~\cite{Ali-Haimoud:2020ozu}:

\begin{equation} \label{sgwb_likelihood}
    -2 \ln{\lc}= {\mathrm{const.}} + \sum_{f}\sum_{AB} \left(\hat{\rc}_A - \frac{\Gamma_{A}^{I}\cdot \bar I}{(4\pi f)^2}  - \frac{\Gamma_{A}^{V}\cdot \bar V}{(4\pi f)^2} \right) \,C^{-1}_{AB}\,\left(\hat{\rc}_B - \frac{\Gamma_{B}\cdot \bar I}{(4\pi f)^2} - \frac{\Gamma_{A}^{V}\cdot \bar V}{(4\pi f)^2}\right).
\end{equation}
In the previous formula, we follow the same notation
of \cite{Cruz:2024svc}, extending 
it by including the circular polarization
contribution $\bar V$. (We refer the
reader to \cite{Ali-Haimoud:2020ozu,Cruz:2024svc} for more information
on this method.)
The sum $\sum_{AB}$ runs over pulsar {\it pairs},
each pair indicated with a capital letter $A=(ab)$.
The dot in eq \eqref{sgwb_likelihood}  indicates a bandwidth-integrated quantity,
centered around the frequency $f$,
hence
\be
\frac{\Gamma_{A}^{I}\cdot \bar I}{(4\pi f)^2} \equiv
\int^{f+\Delta f/2}_{f-\Delta f/2}
\,\frac{\Gamma_{A}^{I} \bar I(\tilde f)}{(4\pi \tilde f)^2}
\,d \tilde f
\ee
and analogously for $V$. The $\hat{\rc}_A$ are
the measured cross-correlations of time residuals. 
The $\Gamma_{A,V}^{I}$ are the PTA response
functions to intensity and circular polarization,
discussed in section \ref{sec_theory}. 
The quantity $C^{-1}_{AB}$ is the inverse of the covariance matrix. We make the hypothesis to work here
in the weak signal limit (more on this in the next section). The covariance matrix  can then be approximated as 
\cite{Ali-Haimoud:2020ozu,Cruz:2024svc}: 

\begin{align}
\label{eq_covm}
    (C)^{-1}_{AB} = \frac{2 \,T_{AB} \Delta f}{{\bf R}_{A}^N\,{\bf R}_{B}^N}\delta_{AB}\,.
\end{align}
with $T_{AA'}$ the time of observation of the four
pulsars $(aba'b')$, and ${\bf R}_{A}^N\simeq \sigma_a^2$ with
$\sigma_a$ the band integrated noise variance
for pulsar $a$.   
 Given a vector $\hat \Theta = {\Theta_i}$
of observable parameters we wish to probe,
the Fisher information criteria  (see e.g. \cite{Tegmark:1996bz})
 allow us to forecast the ability of future
 experiments
 to measure $\Theta_i$.
The best fit correspond to values of $\Theta_i$ 
which makes   vanishing the first derivative
of the likelihood: 
${\partial {\ln {\cal L}}}/{\partial \Theta_i}\,=\,0$.  
 The experimental errors on the parameter
 determination, then, is expressed  in terms of 
 the second-derivatives of the log-likelihood, i.e. the Fisher matrix:

\begin{align}\label{deffim}
        \fc_{ij} = \left\langle -\frac{\partial \ln \lc}{\partial \Theta_i \partial \Theta_j} \right \rangle\,,
\end{align}
where $<\dots>$ indicates we evaluate
the log-likelihood second derivative at the parameter
best fit values. 

\renewcommand{\arraystretch}{1.25}
\begin{table}[h!]
\begin{center}
\begin{tabular}{| c | c | c | c | c | c |}
\hline
\cellcolor[gray]{0.9}$ A_I $ & \cellcolor[gray]{0.9}$ A_V $&\cellcolor[gray]{0.9}$n_I=n_V$&\cellcolor[gray]{0.9}$f_\star $ &\cellcolor[gray]{0.9}$\beta$&\cellcolor[gray]{0.9}$\hat v = (\ell, b)$  \\
\hline
$2.4 \times 10^{-15}$ & $2.4 \times 10^{-15}$ & $ -7/3$ & $1/\mathrm{year}$ & $1.23\times 10^{-3}$ & $(264^{\circ},48^{\circ})$\\
\hline
\end{tabular}
\caption{\it Benchmark values for the parameters to test.\label{tab:valuesa} }
\end{center}	
\end{table}
Armed with this formalism, we investigate in this section how
to use a PTA system to probe to circular polarization for cosmological SGWB, 
making use of the kinematic dipole.
We 
wish to forecast the size of experimental error bars
for a set of GW parameters.
We
assume the  benchmark values  listed in
Table \ref{tab:valuesa}
for
the parameters entering our formulas, 
where the quantities $A_{I,V}$ are introduced in eqs \eqref{defofA}. We examine a situation
characterized by     a large degree of circular polarization $A_V$, of the same order
of the intensity; such a case
can be realized in early universe scenarios where phenomena
of parity violation occur in the gravitational
sector. (For the case of  astrophysical SGWB, see section \ref{sec_astro}.)
 For the intensity parameter $A_I$ we take as reference the  central value measured
  by the NANOGrav collaboration \cite{NANOGrav:2023gor}.
The values characterizing  the kinematic dipole are the   ones obtained by CMB experiments, and
we use  galactic coordinates to express the velocity.
 For the rest of this
section we consider two cases, corresponding to
two different vectors $\hat \Theta$ of parameters
to be tested.

\subsection{Case 1: forecasts on  parameters
$A_V$, $A_I$, and $\beta$}
\label{sub_caseone}

We 
forecast the ability of PTA to measure the 
amplitude $ V_0$ of circular polarization (equivalently $A_V$), 
as well as the SGWB intensity $ I_0$ (equivalently
$A_I$) and kinematic
dipole amplitude $\beta$. 
In the presence  of large values
of $A_V$, a measurement of circular polarization
gives us also  information on the velocity parameter $\beta$ characterizing 
 kinematic anisotropies.
The parameter vector
to be measured is then $\hat \Theta= ( V_0,  A_0, \beta)$. The PTA pulsar set we consider is constituted by the 67 NANOGrav
pulsars, to which we add
 more pulsars located at appropriate
  positions to increase the sensitivity to
  the observables of interest (see section \ref{sec_theory}).

The symmetric Fisher matrix is computed following
 the definition in eq \eqref{deffim}. When evaluated
at a given frequency $f$ it  results (we write each matrix component at leading order in a small $\beta$ expansion)
\bea \label{fisher_IVbeta}
    \fc_{ij}(f) &=&\sum_{A}\frac{2T_{A}\Delta f}{(4\pi f)^4\left(R_A^N\right)^2 }\times
    \nonumber\\
    &&
        \begin{bmatrix}
        \beta^2 \kappa_V^2 \left(\frac{f}{f_\star} \right)^{2 n_V} \left( G_A^{(1)}
        \right)^2 &&\, \,\, \beta \kappa_V  \left(\frac{f}{f_\star} \right)^{ n_V+n_I}  \Gamma_A^{(0)} G_A^{(1)}
        && \,\,\, \beta  \kappa_V  \left(\frac{f}{f_\star} \right)^{ n_V}  G_A^{(1)}[\kappa_{V} {V}_0 G^{(1)}_{A}  + \kappa_{I} {I}_0 \, \Gamma^{(1)}_{A}]\\
        \\
        '' &&  \,\,\,  \left(\frac{f}{f_\star} \right)^{ 2 n_I} \left( \Gamma_A^{(0)}
        \right)^2 &&  \,\,\left(\frac{f}{f_\star} \right)^{  n_I} \Gamma^{(0)}_{A}  [\kappa_{V} {V}_0 G^{(1)}_{A}  + \kappa_{I} {I}_0 \, \Gamma^{(1)}_{A}]
        \\
        ''&&''&& \,\,\,[\kappa_{V} {V}_0 G^{(1)}_{A}  + \kappa_{I} {I}_0 \, \Gamma^{(1)}_{A}]^2
        \end{bmatrix}
        \nonumber\\
\label{fishrea}
\eea
with
\be
\label{defkv}
\kappa_V\,= n_V -1\,\hskip0.5cm,\hskip0.5cm 
\kappa_I\,= n_I -1\,.
\ee
Starting from these equations, 
 the complete Fisher matrix is obtained by summing over the individual frequency bins,
 $\fc_{ij} = \sum_f \fc_{ij}(f)$.  
 We notice that off-diagonal
terms in \eqref{fishrea}
induce correlations
in the measurements
of  the parameters we are
interested in. As a consequence 
a measurement of the kinematic parameter $\beta$ is
facilitated in the presence of large values
of $V_0$ and/or $I_0$, when monitoring pulsars 
located at appropriate positions.

\medskip

\begin{figure}[t!]
    \centering
            \includegraphics[width=0.4\linewidth]{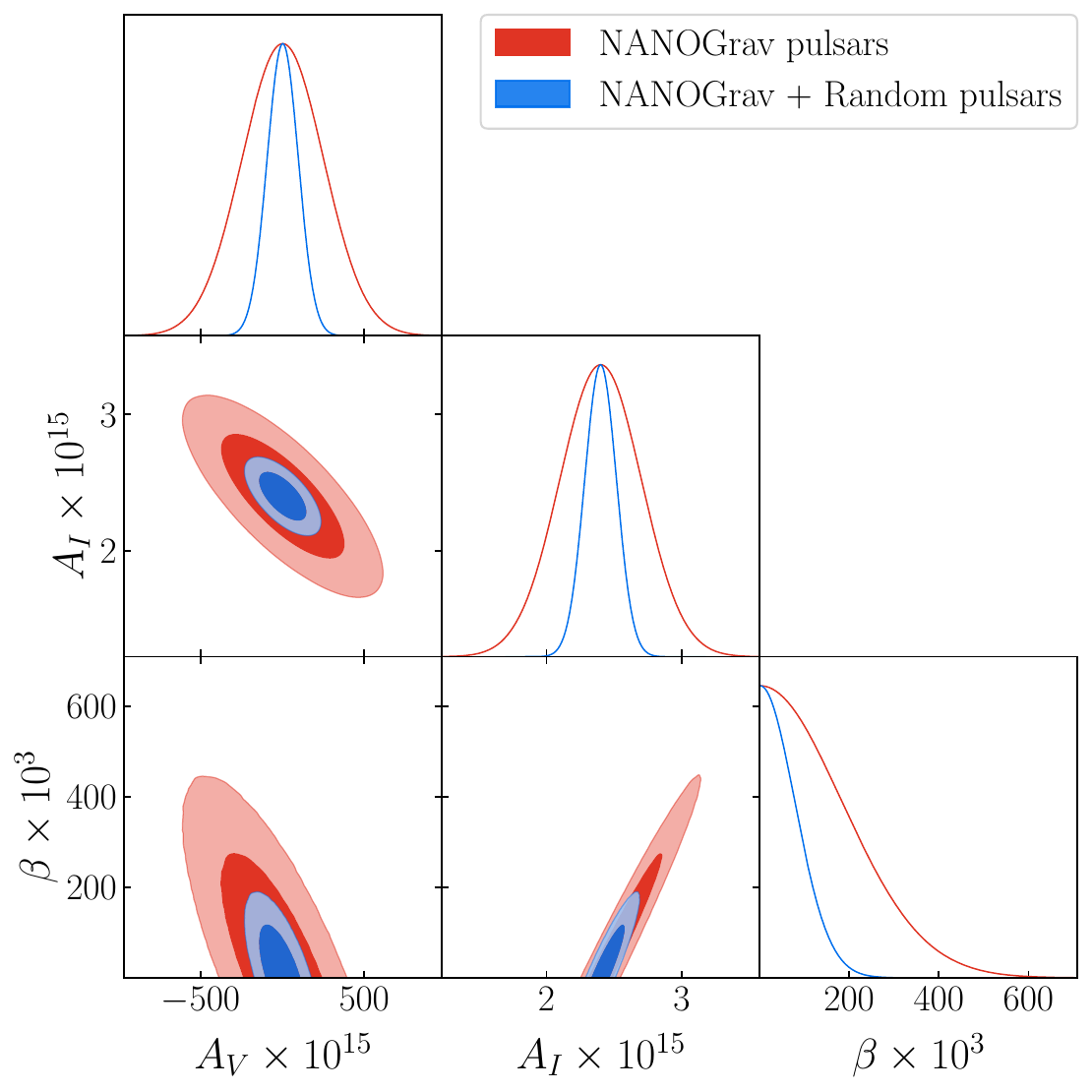}   
               \includegraphics[width=0.4\linewidth]{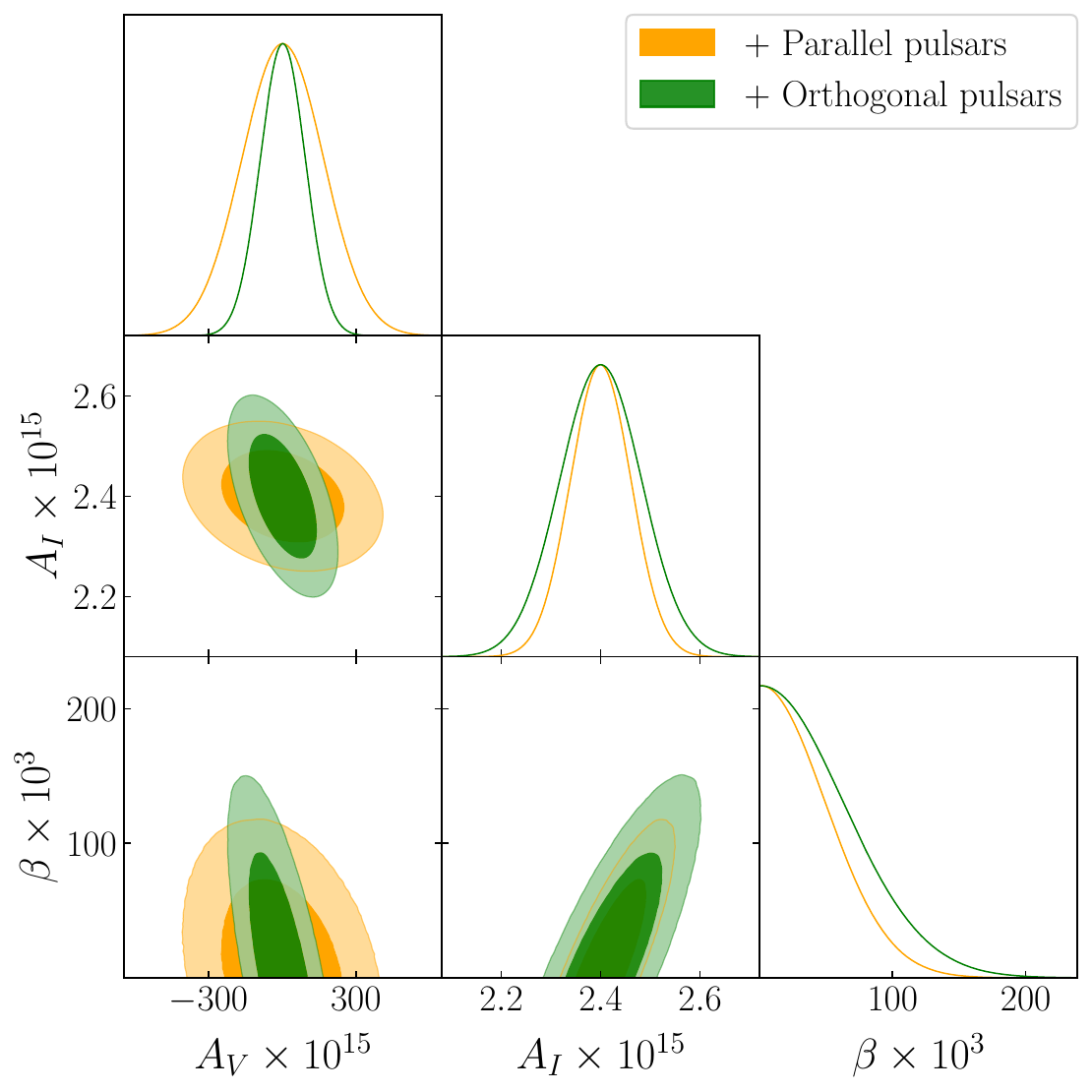}
            \includegraphics[width=0.4\linewidth]{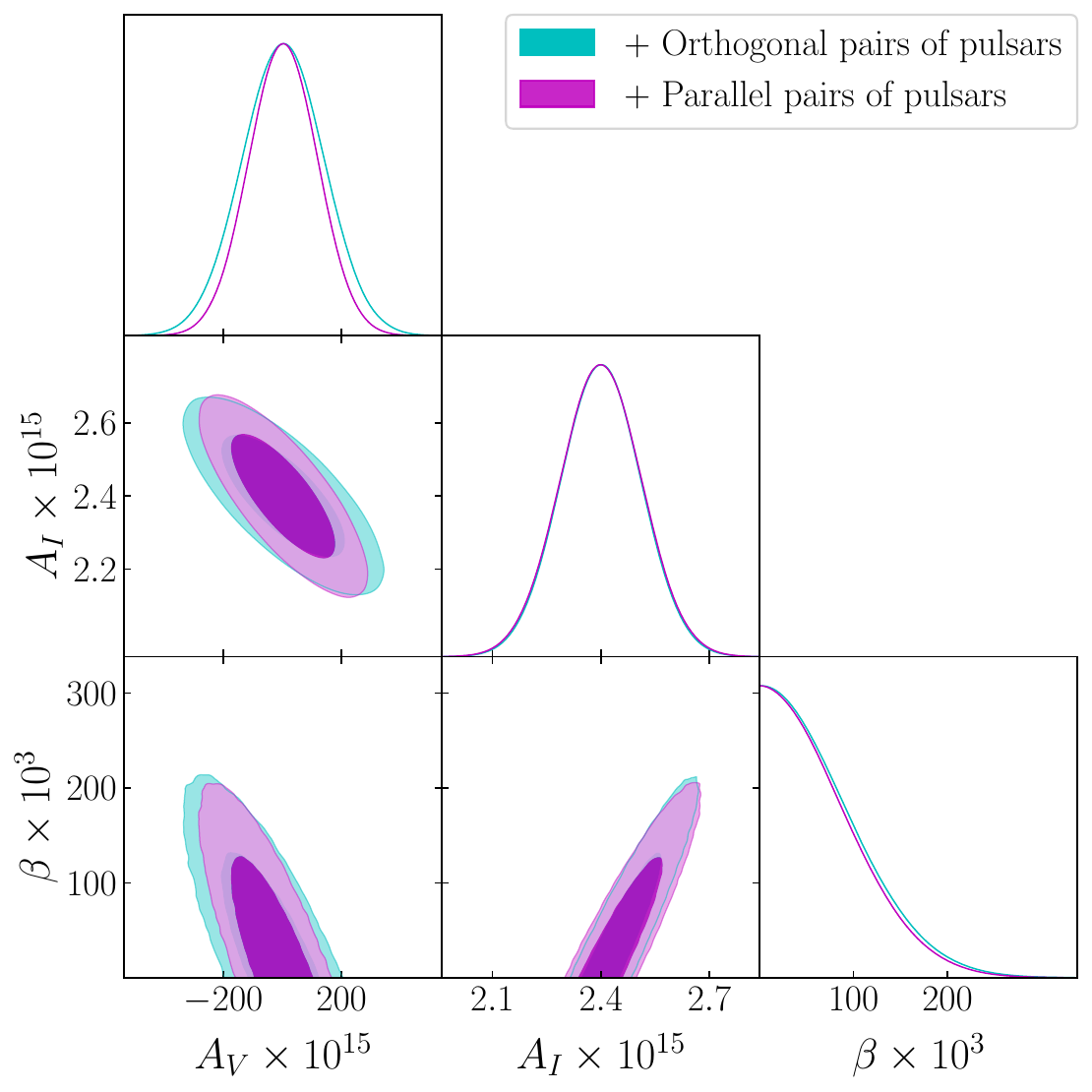}
             \includegraphics[width=0.4\linewidth]{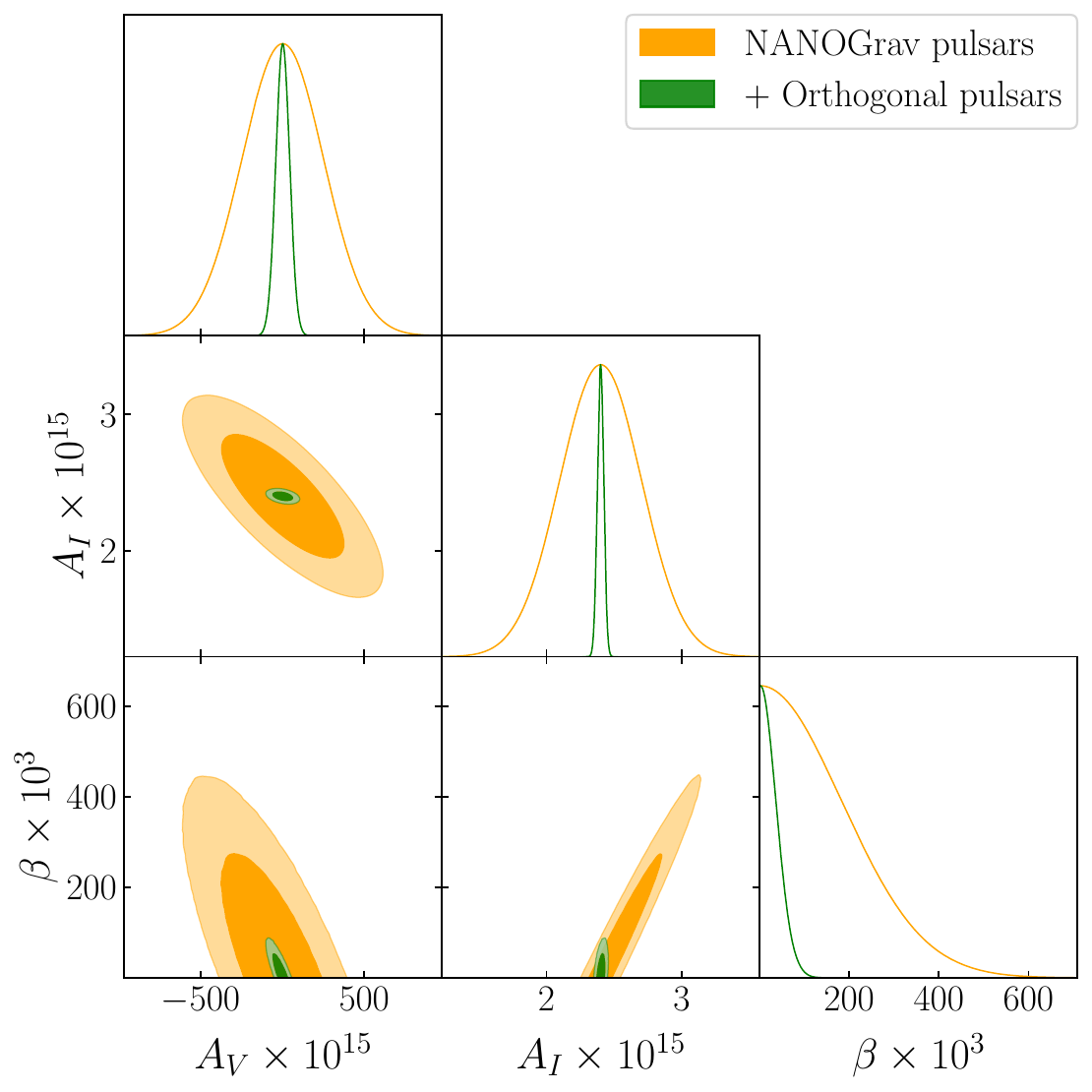}
    \caption{ 
\it Forecasts  for measurements of
 the parameters $A_V$, $A_I$, $\beta$.
  {\bf Upper left panel}: Scenario 1. 
 {\bf Upper right panel}: Scenario 2. {\bf Lower left panel}: Scenario 3. 
{\bf  Lower right panel}: Scenario 4. See main text in section \ref{sub_caseone} for explanations,  and
      Table \ref{tab:valuesb}
      for the error bars. 
    }
    \label{Fig3}
\end{figure}

To exemplify  these statements,
we examine
 four  different  scenarios, 
 with the main aim to quantify
 how the number and location
 of the pulsars to monitor affects the size of the error bars in the measurement of the quantities we are interested in.
 For each scenario, we forecast the size of error bars
 on the parameters
 using the Fisher formalism (see Table \ref{tab:valuesb})
 and we plot them using \texttt{GetDist}~\cite{Lewis:2019xzd}, see Figures \ref{Fig3} and \ref{Fig4}. For each scenario, we add one (or multiple) set(s) of 67
 pulsars, characterized by the same noise
 properties of the 67 
 NANOGrav ones
\cite{the_nanograv_collaboration_2023_8092346}, 
 located at appropriate
 positions in the sky to investigate the effects
 we are more interested on. 

\renewcommand{\arraystretch}{1.25}
\begin{table}[h!]
\begin{center}
\begin{tabular}{| c | c | c | c | c | c | c | c|}
\hline
 {\rm }& \cellcolor[gray]{0.9}$ 1{\texttt{a}} $&\cellcolor[gray]{0.9}$1{\texttt{b}}$&\cellcolor[gray]{0.9}$2{\texttt{a}} $ &\cellcolor[gray]{0.9}$2{\texttt{b}}$&\cellcolor[gray]{0.9}$3{\texttt{a}}$&\cellcolor[gray]{0.9}$3{\texttt{b}}$&\cellcolor[gray]{0.9}$4$  \\
\hline
\cellcolor[gray]{0.9}$ A_V \times 10^{15} = 2.4$ & $\pm250$ & $\pm95$ & $\pm170$ & $\pm92$ & $\pm140$ & $\pm120$ & $\pm42$
\\
\hline
\cellcolor[gray]{0.9}$ A_I \times 10^{15} = 2.4$ & $\pm0.30$ & $\pm0.12$ & $\pm0.061$ & $\pm0.082$ & $\pm0.11$ & $\pm0.11$ & $\pm0.023$
\\
\hline
\cellcolor[gray]{0.9}$\beta \times 10^3 = 1.23$& $\pm180$ & $\pm77$ & $\pm48$ & $\pm61$ & $\pm87$ & $\pm84$ & $\pm35$
\\
\hline
\end{tabular}
\caption{\it Central values and error bars for each scenario considered in  section \ref{sub_caseone}. \label{tab:valuesb} }
\end{center}	
\end{table}

\subsubsection*{$\blacktriangleright$ Scenario 1}

In a first scenario, 
 shown in Fig \ref{Fig3}, upper left panel,
 we consider  the same pulsar set monitored by the NANOGrav collaboration \cite{NANOGrav:2023ctt,NANOGrav:2023hde,the_nanograv_collaboration_2023_8092346,the_nanograv_collaboration_2023_8423265} (red colour, scenario $1{\texttt a}$), and we add to this set 67 randomly located pulsars (blue color, scenario $1{\texttt b}$).  
 Our aim is to understand
 how the addition of pulsars at random positions
 reduces the error bars. In passing from
 scenario $1{\texttt a}$ to $1{\texttt b}$ the errors on $A_V$
 and $\beta$ decrease  by more than a factor of 2
 (see Table \ref{tab:valuesb}), although
 also for scenario $1{\texttt b}$
  they are still large. 
 The measurement of $A_I$,
 which can be obtained by the study of the isotropic
 part of SGWB with no need of detecting anisotropies,
 is very accurate, with a relative error  of order one percent.

\subsubsection*{$\blacktriangleright$ Scenario 2}

In a second scenario, see Fig  \ref{Fig3}, upper right panel,
we instead add pulsars at positions $\hat x$ 
parallel (yellow color, scenario $2{\texttt a}$) or orthogonal (green color, scenario $2{\texttt b}$) to the direction of frame velocity $\hat v$ (we assume for
the new pulsars the same noise curves as NANOGrav pulsars). We learned in section \ref{sec_theory} that when monitoring pulsars located 
with these criteria we
 increase the sensitivity to (respectively) intensity $A_I$ and circular polarization $A_V$  of the SGWB (see \cite{Tasinato:2023zcg,Cruz:2024svc}). In fact, the yellow ellipses in Fig  \ref{Fig3} (upper right panel) demonstrate that measurements of $A_V$ do not
 result 
 more accurate with respect to the case of NANOGrav pulsars only (compare with the upper left panel in the same figure). At the same time,
  the green ellipses   show that, as expected, the accuracy
 in measuring $A_V$ increases
 and the  error halves its size.
 See the error bars in Table \ref{tab:valuesb} for quantitative estimate of the errors. 
 
 \subsubsection*{$\blacktriangleright$ Scenario 3}

In the third scenario, see Fig \ref{Fig3} lower left panel, 
 we  focus on investigating from another viewpoint 
 the sensitivity of the PTA system to circular polarization.
 We start from noticing that 
 eq \eqref{defgv1} indicates   that the sensitivity to $V$  increases when
the condition $|\left[ \hat v\cdot (\hat x_a\times \hat x_b)\right]|=1$ is satisfied. This condition requires
that $\hat v$ is perpendicular to the pulsar
directions (a case already explored in the previous scenario 2), and
additionally that
 $\hat{v}$ is \textit{parallel} to $(\hat x_a\times \hat x_b)$. To analyse this particular case, we
 consider the 67 NANOGrav pulsars, and we 
 select what we call pulsar $\hat{x}_a$ 
  at the position of pulsar B1937+21~\cite{NANOGrav:2023hde}. We then generate
67 extra pulsars to monitor, with positional
vectors $\hat{x}_b$  satisfying the condition $|\left[ \hat v\cdot (\hat x_a\times \hat x_b)\right]|=0$: we call this
scenario $3{\texttt a}$. Similarly, we generate 67 extra pulsars 
with positional vectors satisfying the condition $|\left[ \hat v\cdot (\hat x_a\times \hat x_b)\right]|=1$: this
is scenario $3{\texttt{b}}$. We expect that while 
the sensitivity to $A_V$  increases in scenario $3{\texttt a}$, it is reduced in the scenario $3{\texttt b}$, also with respect
to the original 67 NANOGrav set. 
This behaviour is
confirmed 
by Fig \ref{Fig3}, lower left panel.

 \subsubsection*{$\blacktriangleright$ Scenario 4}

The analysis 
of the previous three scenarios shows that adding just 67 more
pulsars to the NANOGrav data set is not sufficient
to reach the accuracy for an informative measurement
of circular polarization, even if the pulsars are located
to the most convenient positions. In the fourth scenario
we explore
the possibility of  adding { four} extra sets of
67 pulsars each  to the original NANOGrav set (hence, in total, we have 335 pulsars to monitor). The new
pulsars are located orthogonally to the velocity $\hat v$, and each set of 67 pulsars uses the same
noise curves of the 67 NANOGrav pulsars \cite{the_nanograv_collaboration_2023_8092346}. 
The results are represented
in 
Fig \ref{Fig3}, lower right panel, while the error bars
are in Table \ref{tab:valuesb}. We notice that the sensitivity to circular polarization is much
improved
with respect to the previous scenarios (i.e. we gain a factor 8 in sensitivity with respect
to NANOGrav pulsar set). However the error bars are still relatively large. We now proceed to investigate
whether the situation changes by consider
a different set of parameters to constrain.

\subsection{ Case 2: Forecasts
on the parameters $A_V$, $n_V$}
\label{sub_casetwo}

\begin{figure}[t!]
    \centering
\includegraphics[width=0.4\linewidth]{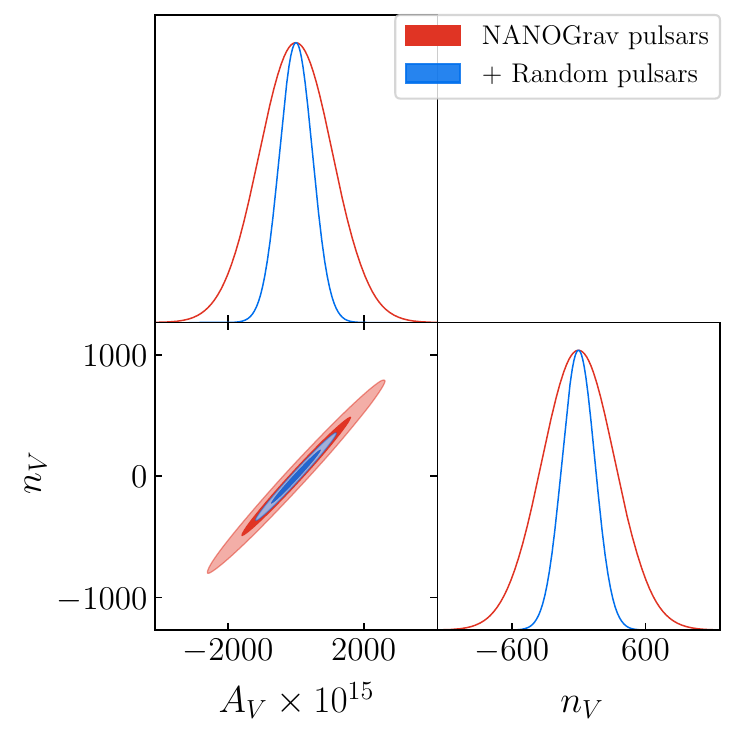}
\includegraphics[width=0.4\linewidth]{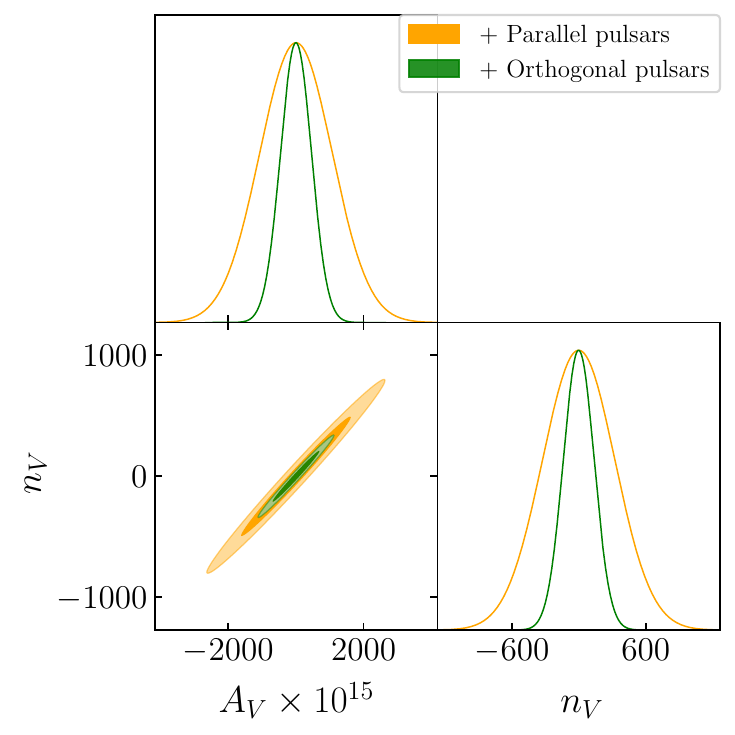}
\includegraphics[width=0.4\linewidth]{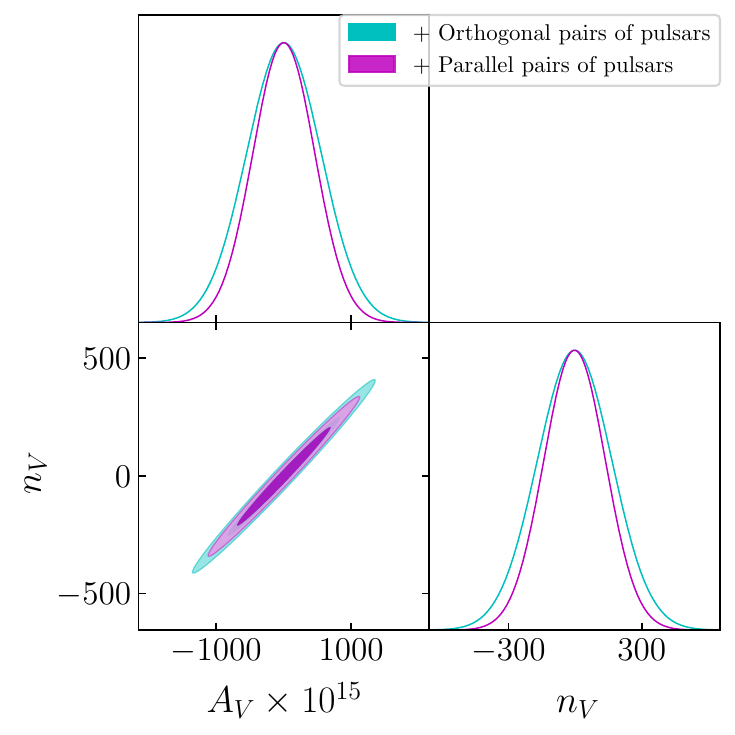}
\includegraphics[width=0.4\linewidth]{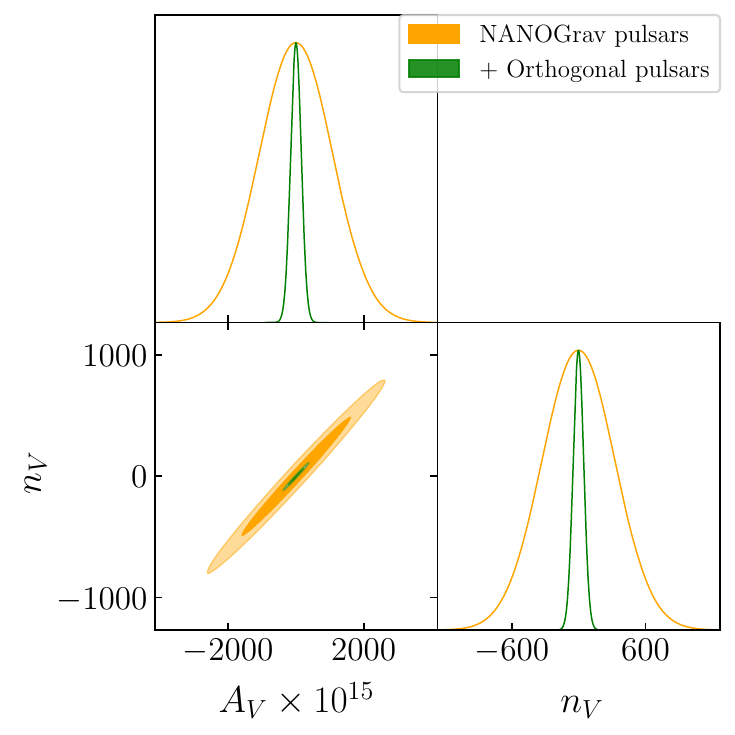}
 \caption{ \it Forecasts for the parameters $A_V$ and $n_V$.
 {\bf Upper left panel}: Scenario 1. 
 {\bf Upper right panel}: Scenario 2. {\bf Lower left panel}: Scenario 3. 
{\bf  Lower right panel}: Scenario 4. See main text in section \ref{sub_casetwo}
  for explanations, and Table \ref{tab:valuesnV} for the corresponding
  error bars.
    }
    \label{Fig4}
\end{figure}
As a second case, we focus on the measurement
of parameters $\vec \Theta = (A_V, n_V)$, assuming
we already have information on the remaining
parameters in the benchmark
set of Table \ref{tab:valuesa}, which enter the likelihood function. Hence we wish
to make use of kinematic anisotropies 
not only to probe the amplitude, but also the slope $n_V$
of the circular polarization spectrum:
see the definition in eq \eqref{intraiav}.
The corresponding
Fisher matrix, evaluated in a Taylor
expansion up order $\beta^2$, results

\bea
    \fc_{ij}(f) &=& \frac{1}{(4\pi f)^4}\sum_{A}\frac{2T_{A} \Delta f (\beta \, G_{A}^{(1)})^2}{\left(R_A^N\right)^2 } \left(\frac{f}{f_{\star}}\right)^{2 n_V}
        \,\begin{bmatrix}
        \kappa_{V}^2 & \kappa_{V} V_0 \left(1 + \kappa_{V} \ln{\left(\frac{f}{f_{\star}}\right)}\right) \\
        \, \kappa_{V} V_0 \left(1 + \kappa_{V} \ln{\left(\frac{f}{f_{\star}}\right)}\right) & \left[V_0 \,\left(1+ \kappa_{V} \ln{\left(\frac{f}{f_{\star}}\right)}\right) \right]^2
        \,
    \end{bmatrix}
    \nonumber\\
    \label{fisherV0nV}
\eea

\renewcommand{\arraystretch}{1.25}
\begin{table}[h!]
\begin{center}
\begin{tabular}{| c | c | c | c | c | c | c | c|}
\hline
 {\rm }& \cellcolor[gray]{0.9}$ 1{\texttt{a}} $&\cellcolor[gray]{0.9}$1{\texttt{b}}$&\cellcolor[gray]{0.9}$2{\texttt{a}} $ &\cellcolor[gray]{0.9}$2{\texttt{b}}$&\cellcolor[gray]{0.9}$3{\texttt{a}}$&\cellcolor[gray]{0.9}$3{\texttt{b}}$&\cellcolor[gray]{0.9}$4$  \\
\hline
\cellcolor[gray]{0.9}$ A_V \times 10^{15} = 2.4$ & $\pm1100$ & $\pm490$ & $\pm1100$ & $\pm460$ & $\pm550$ & $\pm460$ & $\pm160$
\\
\hline
\cellcolor[gray]{0.9}$n_V =-2$& $\pm330$ & $\pm150$ & $\pm330$ & $\pm140$ & $\pm170$ & $\pm140$ & $\pm47$
\\
\hline
\end{tabular}
\caption{\it Central values and error bars for each 
of the scenarios discussed in  section \ref{sub_casetwo}.\label{tab:valuesnV}}
\end{center}	
\end{table}
We consider in this context the same four scenarios
examined in section \ref{sub_caseone}. We represent
our results in Fig \ref{Fig4}, and report 
the corresponding error bar estimate in table \ref{tab:valuesnV}.

In the first scenario, upper left panel of Fig. \ref{Fig4}, we learn that the error bars on $A_V$ and $n_V$ decrease their size (by a factor
one half)
  with  increasing the number
  of pulsars to detect,  as expected. In the
  second scenario, upper right panel of Fig. \ref{Fig4},  we learn  that the error bars are blind to the addition
  of pulsars, as expected from 
  our theory formulas of section \ref{sec_theory}.
 Instead, the error bar sizes decrease considerably adding 
 pulsars located orthogonally to the velocity among frames
 (see lower left panel of  Fig \ref{Fig4})
 . 
 The same behaviour is confirmed in the 
   fourth scenario plotted in the lower
   right panel of Fig \ref{Fig4}. 
   
Still, the error bars in the parameters are quite large, even for scenario $4$.To reduce the  error bars of our forecasts to a level acceptable for ensuring detection, we discuss in section \ref{sec_fut} a method to analytically explore more futuristic set ups able to considerably
enhance the sensitivity to circular polarization, by increasing
the number of pulsars to be monitored.

\section{Futuristic set up: isotropically distributed pulsars}
\label{sec_fut}

We learned in the previous section that adding  pulsars
at specific positions certainly improves
the detectability prospects of circular polarization $V$, but not enough to ensure
detection. In this
section, following ideas first developed in \cite{Ali-Haimoud:2020ozu}, we focus
on an idealized scenario which  simplifies our
calculations
and allow for analytical estimates
of the expected accuracy obtained with future
measurements.

We 
consider a futuristic setup of PTA monitoring a  a large number of pulsars, say $N_{\rm psr}$
in the order of thousands, and we take 
eq \eqref{sgwb_likelihood} for our likelihood.
For
developing our analytical arguments, we make the following
hypothesis:
\begin{itemize}
\item We assume the pulsars are  isotropically distributed across the sky, they
have the same properties, and they are monitored for the same amount of time.
\item For the intrinsic noise common to all pulsars  we 
select red noise parameters which lie slightly towards the higher end of those observed in the NANOGrav dataset~\cite{NANOGrav:2023hde}, which allows
us to work in the weak-signal regime with a diagonal covariance matrix, as
in eq~\eqref{eq_covm}. 
\end{itemize}
These simplifying assumptions allow us to approach 
the problem analytically, using the Fisher formalism of~\cite{Ali-Haimoud:2020ozu}. 

We assume that the properties of the SGWB intensity (including the kinematic dipole) are already known, thus we fix $\beta,\,n_I,\,A_I$ to the values used in the previous section, see Table \ref{tab:valuesa}.\footnote{Our choice of fixing the intensity parameters for the Fisher forecasts -- instead of considering them as additional parameters to be measured simultaneously -- will be justified in what comes next.} 

Analogously to 
what done in the literature for
the intensity of the SGWB~\cite{Ali-Haimoud:2020ozu,Cruz:2024svc}, in our limit of a large number of identical, isotropically distributed pulsars we obtain the following Fisher matrix for the circular polarization parameters, evaluated
at fixed frequency $f$
\label{sec:LargeN}
\label{sec:Fisher_ideal}
\begin{align}
    \fc_{ij}(f) = \frac{2T\Delta f}{(4\pi f \sigma)^4}N_{\rm pair}\times \frac{1}{N_{\rm pair}}\sum_{pq} \gamma_{pq}^V \cdot \frac{\partial V}{\partial \Theta_i} \gamma_{pq}^V \cdot \frac{\partial V}{\partial \Theta_j}\,, 
\end{align}
where $V$ is given in eq \eqref{intraiav}, and $\gamma^V_{pq}$ in eq \eqref{def_gav}. 
This Fisher component should then be summed over  frequencies, 
as in the previous section. 
 Following~\cite{Ali-Haimoud:2020ozu} (which focuses on   the intensity in the dense pulsar limit), we can convert the sum
into  an integral, and write 
\begin{align}
    \frac{1}{N_{\rm pair}} \sum_{pq} \gamma_{pq}^V \cdot \frac{\partial V}{\partial \Theta_i} \gamma_{pq}^V\cdot \frac{\partial V}{\partial \Theta_j} = \int \frac{d^2 p}{4\pi} \frac{d^2 q}{4\pi} \int \frac{d^2 n}{4\pi} \frac{d^2 n'}{4\pi} \gamma_{pq}^V(\hn)\gamma_{pq}^V(\hn') \frac{\partial V(\hn)}{\partial \Theta_i}\frac{\partial V(\hn')}{\partial \Theta_j}\,.
\end{align}
 We then consider the quantity 
\begin{align}
    \lim_{N_{\rm psr}\to \infty}  \frac{1}{N_{\rm pair}} \sum_{pq} \gamma_{pq}^V(\hn) \,\gamma_{pq}^V(\hn') &= \int \frac{d^2 \hp}{4\pi} \frac{d^2 \hq}{4\pi} \gamma_{pq}^V(\hn) \,\gamma_{pq}^V(\hn') \\
    &\equiv \mathcal{F}^V_{\infty}(\hn \cdot \hn')\,,
\end{align}
for which  for the first time
 we obtain the following novel analytic expression~\footnote{This result
can be in principle obtained following
the methods of Appendix C of \cite{Ali-Haimoud:2020ozu}. We found
it easier to carry the integrals in the complex plane
and use Cauchy theorem, applying in this context the procedure proposed
in \cite{Jenet:2014bea} for computing the Hellings-Downs curve. Please contact the authors for obtaining the Mathematica notebooks
leading to eq \eqref{eq_nfcp}.
\label{footexp}}
\begin{eqnarray}
\fc_{\infty}^V (\hn \cdot \hn')&=&
\frac{8\left(
18-17 \chi+2 \chi^2+\chi^3
\right)}{9 (1+\chi)^2}
+ 
\frac{8 \left(
17-25 \chi+7 \chi^2+\chi^3
\right)
}{3 (1+\chi)^3}
\ln{\left( \frac{1-\chi}{2}\right)}
\nonumber
\\
&&+\frac{32 (1-\chi)^2}{ (1+\chi)^4}\,\ln^2{\left( \frac{1-\chi}{2}\right)}
\label{eq_nfcp}\,,
\end{eqnarray}
where $\chi\equiv \hn\cdot\hn'$.  Via the approach explained in
footnote \ref{footexp} one can  show that the quantity $\fc^{IV}_{\infty}=0$ in the weak signal limit, implying that in the limit of a large number of identical pulsars distributed isotropically across the sky, there is no `leakage' of the intensity map into the circular polarization map. This justifies our choice of assuming the intensity is known when performing the Fisher forecasts for circular polarization.

\begin{figure}[ht]
    \centering
    \includegraphics[width=0.98\linewidth]{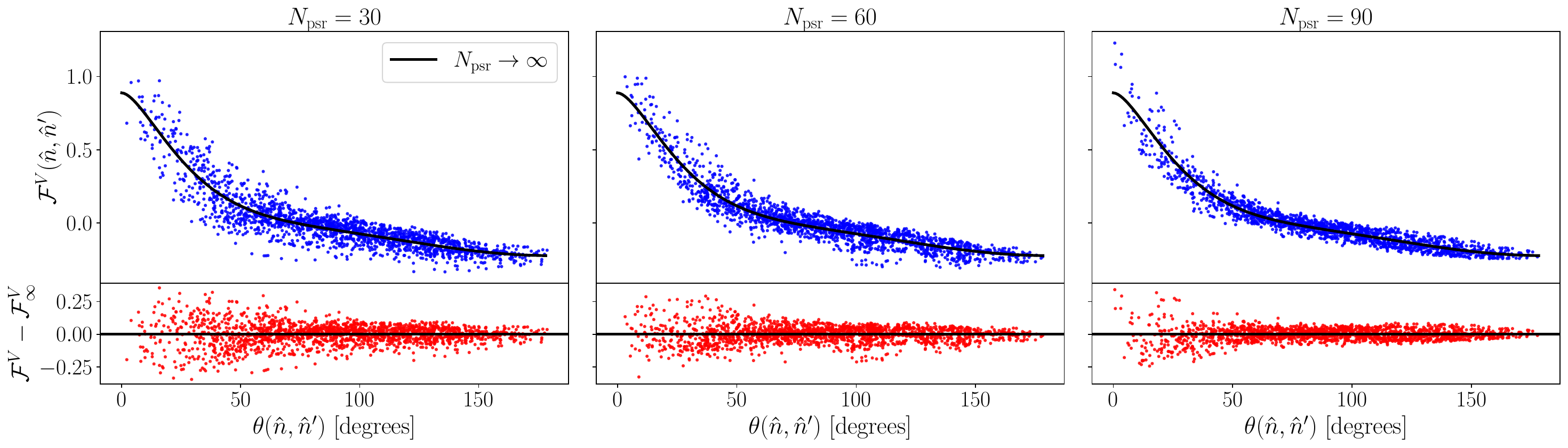}
    \caption{\textit{A plot of $\mathcal{F}^V(\hn\cdot\hn')$ in a realistic case with $N_{\rm psr}$ pulsars distributed isotropically across the sky, as a function of the angle between the sky directions $\hn$ and $\hn'$. The black curve shows the exact result in the limit $N_{\rm psr}\to \infty$. Notice that as $N_{\rm psrs}$ increases, the scattering  of finite pulsar results with respect to  the  limit of eq \eqref{eq_nfcp} decreases.}}
    \label{fig:FV_inf}
\end{figure}

\begin{figure}
    \centering
    \includegraphics[width=0.5\linewidth]{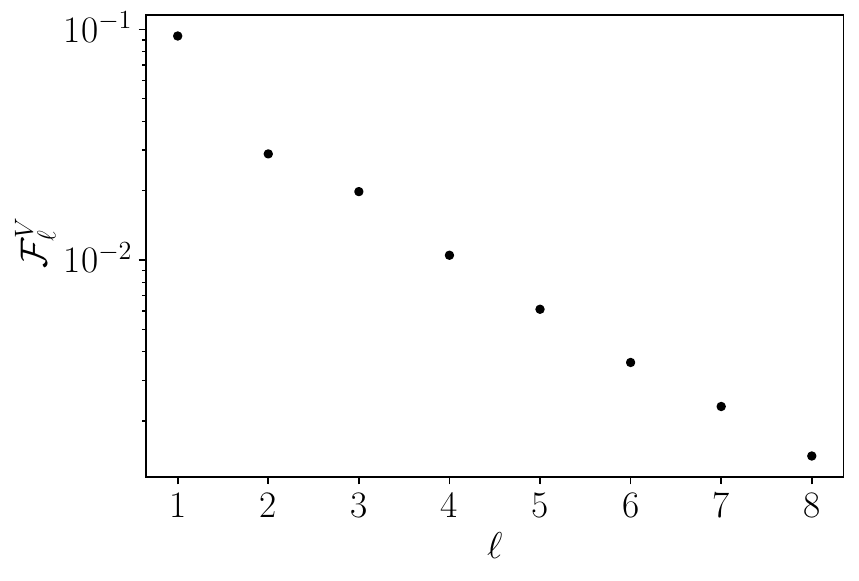}
    \caption{\it The spherical harmonic coefficients $\fc^V_\ell$ relative to eq  \eqref{eq:sph_VF}.}
    \label{fig:FV_ell}
\end{figure}

 Having calculated $\fc^V_{\infty}$ (plotted in \cref{fig:FV_inf}), we can expand it in spherical harmonics (we are justified by the fact that we assume the PTA system to be isotropically distributed in the sky):
 \begin{align}
 \label{eq:sph_VF}
    \fc^V_{\infty}(\hn \cdot \hn') &= 4\pi \sum_{\ell m} \fc^V_{\ell}\,Y_{\ell m}(\hn)Y_{\ell m}^*(\hn')
\end{align}
The spherical harmonic coefficients of $\fc^V_{\infty}$ in eq \eqref{eq:sph_VF} read $\fc^V_1 = 0.094,\, \fc^V_2 = 0.029,\, \fc^V_3 = 0.02$. {Interestingly, in this idealised full sky limit we find $\fc_1^V \approx 5 \fc_1^I$ (see~\cite{Ali-Haimoud:2020ozu,Cruz:2024svc} for $\fc^I$), which suggests that for sizable polarization, it may be easier to detect the  polarization dipole as compared to the dipole relative to the SGWB intensity.} The $\fc_\ell^V$ are plotted in \cref{fig:FV_ell}. Note that -- as expected -- there is no $\ell=0$ monopole response, since PTA experiments are blind to the circular polarization monopole.
We do the same for the $V$-mode map, where we consider only the dipole term,
the dominant one in a small $\beta$
expansion 
\begin{align}
    \label{eq:sph_VF2}
    V(f,\hn)&=  \bar{V}(f) \,\frac{4\pi}{3}\beta (n_V-1)\sum_{m}Y_{1m}(\hn)Y_{1m}^* (\hat v)
\,.
\end{align}
Substituting the above expressions into our formula for the Fisher matrix  evaluated
at a given frequency $f$, we find
\begin{align}
    \fc_{ij}(f) = \frac{2T\Delta f }{(4\pi f \sigma)^4}N_{\rm pair}
    \times  \frac{4\pi\fc^V_1}{9}\sum_{m}\frac{\partial V_{1m}}{\partial \Theta_i}\frac{\partial V_{1m}^{*}}{\partial \Theta_j}
    \,,
\end{align}
where
\begin{align}
    V_{1m} \equiv  \bar V \,\beta\,(n_V-1) \,Y_{1m}(\hv{v})
    \,,
\end{align}
We now 
forecast the prospects for future PTA experiments
 to measure the  circular polarization amplitude $ V_0$  and the circular polarization tilt $n_V$ introduced in eq \eqref{intraiav}. 
 We find, at leading order in a $\beta$ expansion, 
\begin{align}
    \label{eq:Fisher_V_nV}
    \fc_{ij}(f) =&   \frac{\fc^V_1}{3}\left(\frac{f}{f_{\star}}\right)^{2 n_V}
    \begin{bmatrix}
        \kappa^2\beta^2  &  {V}_0\kappa\beta^2 \left[1+ \kappa\log ({f}/{f_{\star}}\right] \\
         {V}_0\kappa\beta^2\left[1+ \kappa\log ({f}/{f_{\star}})\right]  &{V}_0^2\beta^2\left[1+\log ({f}/{f_{\star}}) (n_V-1)\right]^2 \nonumber \\
    \end{bmatrix} \\ &\times \frac{2T\Delta f }{(4\pi f \sigma)^4}N_{\rm pair} + \mathcal{O}(\beta^3)
    \,.
\end{align}
Once again, the full Fisher matrix will then be obtained by summing over the individual frequency bins. 

We start with the  simplest case, and  we assume the polarization power law index is the same as the intensity (taken to be $n_V=n_I = -2/3)$: we  then fix this quantity. We assume dipole amplitude and direction to be the same as inferred from the CMB (see Table \ref{tab:valuesa}) and we first focus on the estimate of $|\ev| \equiv (A_V/A_I)^2$  only. In this case, the Fisher matrix reduces to
\begin{align}
    \fc_{\ev} &= \sum_f \frac{2T\Delta f }{(4\pi f \sigma)^4}N_{\rm pair}\frac{\fc^V_1}{3}\left(\frac{f}{f_{\star}}\right)^{2 n_V} {V}_0^2\kappa^2  \beta^2 \nonumber \\
    & = \sum_f \frac{2T\Delta f }{(4\pi f \sigma)^4}N_{\rm pair}\frac{\fc^V_1}{3}\left(\frac{f}{f_{\star}}\right)^{2 n_V}\kappa^2 {V}_0^2\beta^2 \times \frac{\fc_0 \bar{I}^2}{\fc_0 \bar{I}^2} \\
    & = \SNR^2_{I} \times \epsilon^2_V \kappa^2 \beta^2 \frac{\fc^V_1}{3\fc_0} \nonumber
\end{align}
where
\begin{align}
    \SNR_{\rm I}^2 =  \sum_{f}\frac{2T\Delta f }{(4\pi f \sigma)^4}N_{\rm pair} \left(\frac{f}{f_{\star} }\right)^{2 n_V} \bar{I}^2 \fc_0\, 
        \,.
\end{align}
Our PTA setup has the same configuration we used for the intensity forecasts in~\cite{Cruz:2024svc}. We take $T_{\rm obs}=20\,\mathrm{years}$ for the pulsar white noise; we fix \mbox{$T_{\rm cad} = \mathrm{year}/20$} as the cadence, and $\Delta t_{\rm rms} =100\,\mathrm{ns}$ residuals~\cite{Hazboun:2019vhv}. For the pulsar red-noise, we fix $A_{\rm RN}=2\times 10^{-15}$ and $\alpha_{\rm RN}=2/3$ which are input into \texttt{Hasasia}~\cite{Hazboun2019} to obtain the noise curves.   Our frequency bins correspond to $\Delta f = 1/T_{\rm obs}$ as the bin width and $f\in [1/T_{\rm obs}, 20/T_{\rm obs}]$ as the bin centers.\\

 The estimated error in the measurement of $\epsilon_V$ is given by,
\begin{align}
    \Delta \epsilon_V = \sqrt{1/\fc_{\epsilon_V}} = \left[\frac{3\fc_0}{\epsilon_V^2 \kappa^2 \beta^2 \fc^V_1 \SNR^2_{\bar{I}}}\right]^{1/2}\,. 
\end{align}
\begin{figure}[ht]
    \centering
    \includegraphics[width=0.5\linewidth]{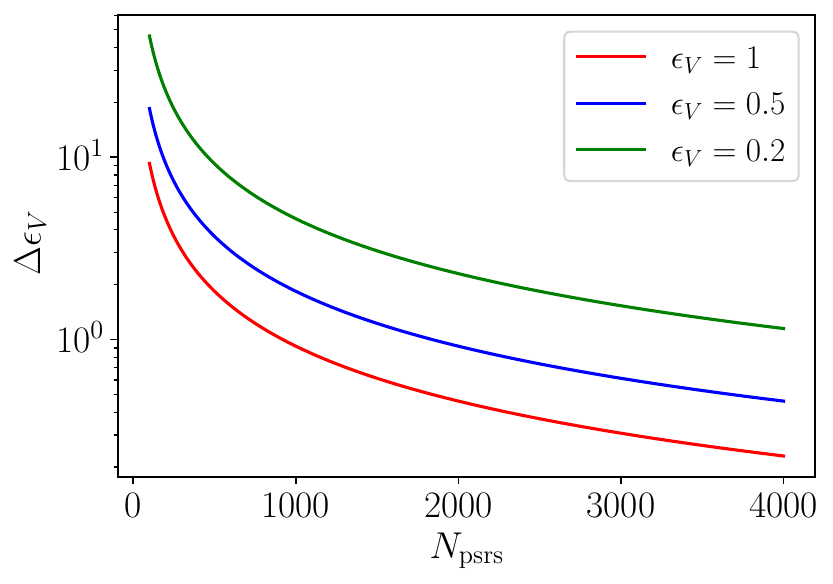}
    \caption{\textit{Error in measurement of the re-scaled circular polarization parameter $\epsilon_V$ as a function of $N_{\rm psr}$ for different values of $\ev$, with fixed spectral tilt $n_V=n_I=-7/3$.}}
    \label{fig:error_amplitude_only}
\end{figure}
This result is plotted in Fig \ref{fig:error_amplitude_only} as a function of $N_{\rm psrs}$ for different values of $\epsilon_V$. The figure demonstrates  that, under the conditions we are considering, we need at least {\bf 1000 pulsars} to reach an accuracy sufficient to claim a detection of circular polarization at confidence level greater than 95\% for $\ev \geq 0.5$. Definitely, if such futuristic set ups can be achieved, as for example in
SKA-type experiments,
tests of circular polarization with PTA can be very accurate. On the other hand, for $\ev<0.5$, with such a PTA configuration $\Delta \ev /\ev$ still remains quite large. 
We stress that in this section, differently from section
 \ref{sec_present}, we also adopt the same noise curve for each of the pulsars. This choice, along with the fact that we only focus on the polarization amplitude here, reduces
 considerably the error bars and improves
 prospects of detection.

\smallskip

\begin{figure}[ht]
    \centering
    \includegraphics[width=0.45\linewidth]{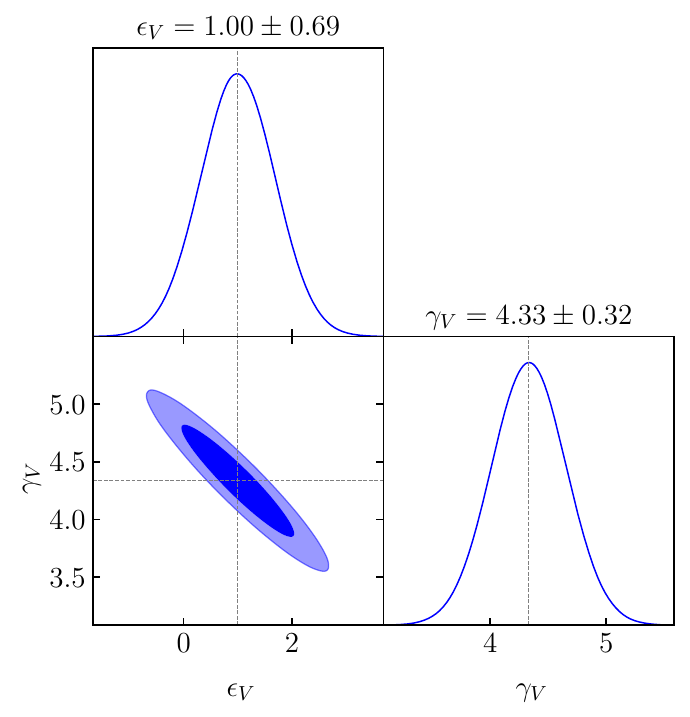}
    \includegraphics[width=0.45\linewidth]{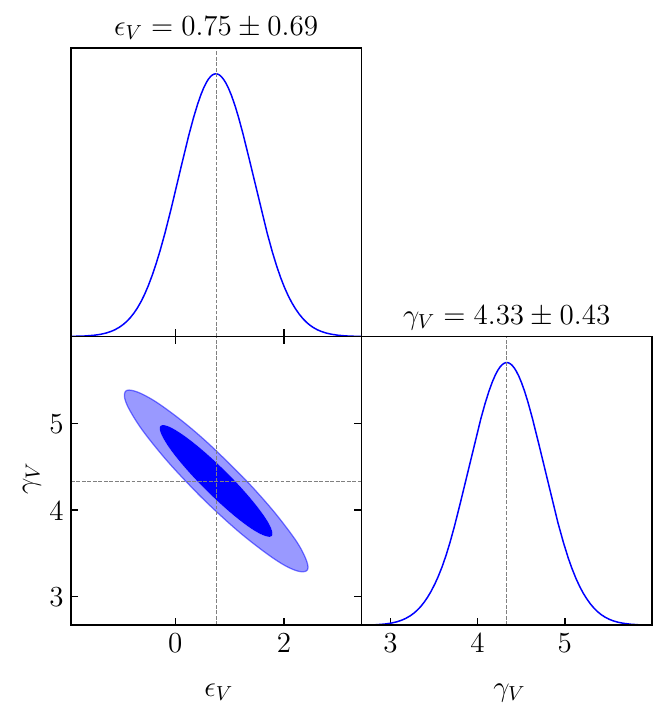}
    \caption{\textit{Fisher forecasts for the re-scaled amplitude and the spectral tilt. The number of pulsars is taken to be $N_{\rm psr}=4000$ for both cases.}}
    \label{fig:Fisher_eV_ideal}
\end{figure}

As a second case,
we now consider both the amplitude and spectral tilt of circular polarization as parameters to be estimated from the data. 
We express the spectral tilt in terms of the combination $$\gamma_V=2-n_V\,.$$
The resulting Fisher matrix is the same as in the first case, up to re-scalings arising from the change of variables ${V}_0$ to
$\ev$. The fiducial signal parameters are the same as in Case I and we provide forecasts for two different values of $\ev = 0.75,1$. These are plotted in \cref{fig:Fisher_eV_ideal}.
We learn that a maximal level of polarization may be detected with such a PTA whereas for the $\ev=0.75$ case, the error bars are large $\Delta\ev/\ev \simeq 1$.  Notice that when the spectral slope is an additional parameter to be estimated, for the same PTA setup as in~\cite{Cruz:2024svc}, the relative error on the amplitude is larger. 
The main reason being that the circular polarization monopole is not detectable with PTAs, thus if one does not make the assumption that $n_V = n_I$, both the polarization amplitude and spectral tilt have to be estimated together from the anisotropies. This  degrades the sensitivity to the amplitude owing to the strong degeneracy between the spectral tilt and the polarization dipole amplitude, as can be seen from~\eqref{GammaABV}. In fact, the correlation coefficient between $\ev$ and $n_V$, which is given by $ C_{10}/\sqrt{C_{11}C_{00}}=0.94$, is quite large for the $\epsilon_V=1$ case, indicating that any  estimates of the two parameters will be strongly correlated. For
this reason,  the error bars on $\ev$ are larger compared to those observed in~\cref{fig:error_amplitude_only}. 

 Note that decreasing the noise level will not directly translate into a reduction  in the error bars on $\beta$ and $n_V$, as one would expect from the form of the Fisher matrix. The reason being that decreases in the noise level will lead to the PTA entering an intermediate or strong-signal regime where the signal to noise ratio only grows as $\SNR \propto N_{\rm psr}\sqrt{T_{\rm obs}}$~\cite{Siemens:2013zla}. Indeed, some of the pulsars observed by current PTAs are already in the intermediate signal regime and it is likely that future PTAs will end up being in the strong signal regime. Thus, increasing the number of pulsars observed remains the best bet for more accurate measurement of circular polarization.

This concludes our analysis of
cosmological SGWB. In the next
section, we will learn that more
optimistic prospects of detection
are achieved for SGWB of astrophysical
origin. 

\section{Astrophysical SGWB Anisotropies and Circular Polarization}
\label{sec_astro}

We now consider the case of an astrophysical
SGWB.  The astrophysical
background detectable at PTA frequencies  is characterized
by significant intrinsic anisotropies, with a relative amplitude
of order $\mathcal{O}(10^{-1}\text{--}10^{-2})$ with respect
to the isotropic background (see~\cite{NANOGrav:2023tcn} and references therein).  We  estimate how a measurement of such intrinsic anisotropies can provide information on the SGWB circular polarization.
We will learn that in this case the 
prospects of detection  of this quantity are much more
promising as compared to a cosmological background.

A net circular polarization characterizes the SGWB from  astrophysical
SGWB sources, being 
 generated due to fluctuations in the GW source properties, and the inclination angles of black hole binaries with respect to the line of sight.
  A quantitative estimate of the net circular polarization due to these effects, relevant for the PTA range was performed in ref.~\cite{ValbusaDallArmi:2023ydl,Sato-Polito:2023spo} with the result that the circular polarization anisotropies due to such effects have an amplitude quite close to the corresponding intensity anisotropies \footnote{
These estimates of the intensity and polarization anisotropies are based on shot-noise models for the SMBHB and    
ignore the possible effects of clustering.   
The circular polarization is generated due to random inclinations of binary orbits with respect to the line-of-sight, so if the inclinations of sources are uncorrelated with the locations, i.e. clustering does not affect inclinations then the relative ratio $V/I$ should still be the same.}, i.e. $C_{\ell}^V \approx C_{\ell}^I$, with the same frequency and multipole slope as the intensity. 

In this section we calculate the expected SNR with which astrophysical SGWB circular polarization will be detected in future datasets, working
in the same idealized
framework of section \ref{sec_fut}. We assume that the angular power spectra of circular polarization $C_{\ell }^V (f)$ scale with $\ell$ and $f$ in the same manner as the intensity, i.e.
\begin{align}
    C_{\ell }^V (f) = \epsilon_V^2 C_{\ell  }(f)\,,
    \label{eq:agwb_Clv}
\end{align}
with $|\ev| = (A_V/A_I)^2 <1$. We assume that the $C_{\ell}$ are independent of $\ell$, as found in~\cite{Sato-Polito:2023spo,NANOGrav:2023tcn}. We  calculate the minimum amplitude $|\epsilon_V|$ allowing for a detection of circular polarization, for which we assume a SNR threshold $\SNR^V = 5$. The SNR (at a given frequency) can be expressed as~\cite{Ali-Haimoud:2020ozu} 
\begin{align}
    [\SNR_f^V]^2 = V(f, \hn) \cdot \fc_f(\hn,\hn') \cdot V(f, \hn')
    \,.
\end{align}
Expanding the the polarization map and the Fisher matrix in spherical harmonics,
\begin{align}
    V(f,\hn) = \sum_{\ell m} v_{\ell m}Y_{\ell m}(\hn),\quad \qev{v_{\ell m} v^{*}_{\ell' m'}} \equiv C_{\ell}^V(f)\delta_{\ell \ell'}\delta_{m m'}\,,
\end{align}
 the expected value of the $\SNR^2$ becomes
\begin{align}
    \label{eq:SNR_V_AGWB}
    [\SNR_f^V]^2 =\frac{2T\Delta f}{(4\pi f \sigma)^4}N_{\rm pair} \sum_{\ell>0}^{\ell_{\rm max}} (2\ell+1) C_{\ell}^V(f) \frac{\fc^V_{\ell}}{4\pi}\,.
\end{align} 
The total SNR is obtained by summing over frequencies
\begin{align}
    \label{eq:SNR_V}
    [\SNR^V_{\rm tot}]^2 = \sum_f [\SNR_f^V]^2 \,.
\end{align}
\begin{figure}
    \centering
    \includegraphics[width=0.9\linewidth]{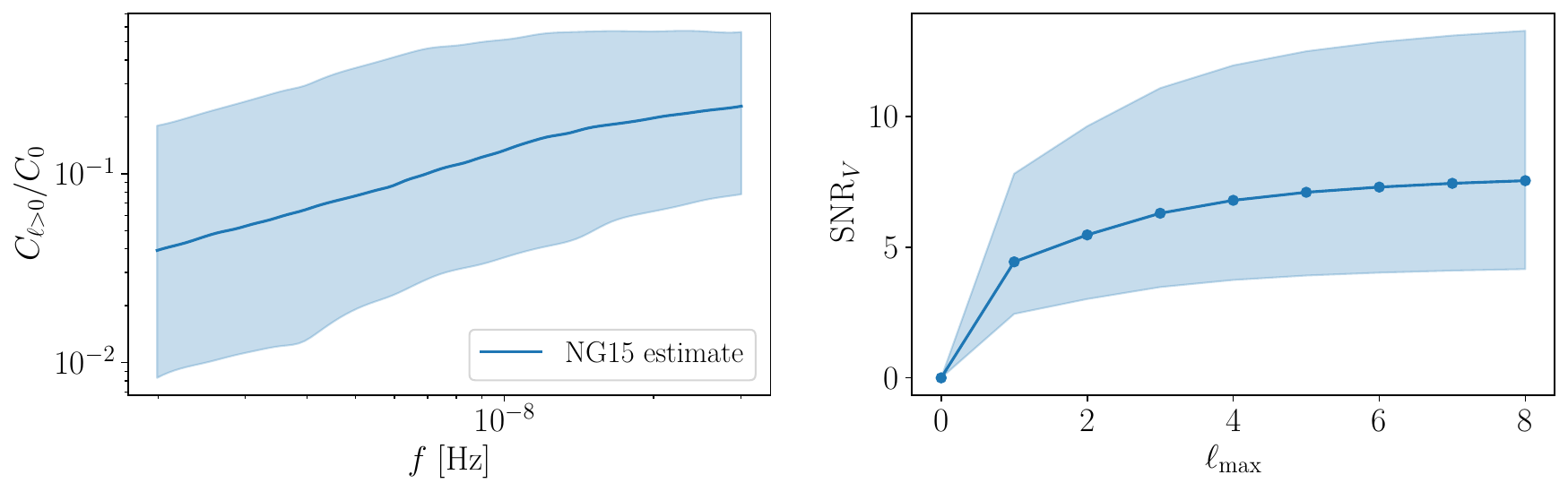}
    \caption{\textit{Left: Estimate of SMBHB intensity anisotropy from NG15, taken from fig.~11 of \cite{NANOGrav:2023tcn}, the shaded region denotes the $68\%$ C.L of the estimate. 
    The $C_\ell$ are normalized to the monopole of the intensity, $C_0$. Right: the circular polarization SNR as a function of $\ell_{\rm max}$ for $\epsilon_V=1$ and $N_{\rm psr}=150$.}}
    \label{fig:SNR_V}
\end{figure}

Our results are summarized in  Fig. \ref{fig:SNR_V}, where we plot the SNR of~\cref{eq:SNR_V} as a function of the maximum multipole $\ell_{\rm max}$, for $\ev=1$. In this case of maximal circular polarization of the AGWB anisotropies, we find that even {\bf 150 pulsars} with the selected noise properties may be enough to detect the astrophysical SGWB circular polarization. The much smaller number of pulsars required with respect
to cosmological sources (recall
section \ref{sec_fut}) is  a consequence of the expected astrophysical SGWB anisotropy being several orders of magnitude larger than the kinematic dipole anisotropy. In the weak signal limit, the corresponding  SNR  scales with the polarization amplitude and number of pulsars as (see~\cref{eq:SNR_V_AGWB})
\begin{align}
    \SNR^V \propto N_{\rm psr} \ev\,.
\end{align}

Additionally, we can infer that for the PTA setup with many more pulsars to be monitored, say  4000 pulsars as considered in the previous section on
cosmological sources, the minimum detectable level of circular polarization is considerably lowered, at the level  $\ev \approx 0.05$. 
The SNR saturates fairly quickly increasing $\ell$, at around $\ell_{\rm max}=8$, which can be understood from the fact that the coefficients $\fc_{\ell}$ which represent the PTA sensitivity to circular polarization decrease rapidly with $\ell$ (see \cref{fig:FV_ell}).

Overall, our results suggest that if the astrophysical SGWB is characterized by a sizeable level of circular polarization, as found in~\cite{ValbusaDallArmi:2023ydl,Sato-Polito:2023spo}, then even near future datasets may be able to set tight constraints on circular polarization. This is especially true since the results of~\cite{NANOGrav:2023tcn} suggest that that the astrophysical  SGWB intensity anisotropy may soon be detectable. Even for much smaller levels of SGWB circular polarization, the prospect of detection with an SKA-level experiment is promising. For realistic near future datasets however, we also need to account for the mixing between $I$ and $V$ anisotropy due to the non-isotropic distribution of a finite number of non-identical pulsars (see eq.~\eqref{fisher_IVbeta}). We plan to explore the joint estimation of the $I$ and $V$ anisotropies in future work. Furthermore, note that at present there is some uncertainty in the expected level of astrophysical SGWB intensity anisotropies, so these results might change as our models of the SGWB anisotropy become more and more sophisticated.

\section{Conclusions}
\label{sec_concl}

The polarization content of the SGWB is a key observable that may be used to characterise the background and possibly discriminate between astrophysical and cosmological origin of the signal observed by PTA collaborations. For astrophysical SGWB, net circular polarization may be generated due to fluctuations in the source properties whereas for cosmological sources this would indicate the presence of parity violating physics in the early universe. The detection of circular polarization is made difficult by  the fact that PTAs are blind to its monopole, thus we need to observe anisotropies, which are smaller in amplitude  and more challenging to detect. 

In this paper we have studied the prospects of detecting circular polarization with near and far future PTA experiments, focusing on both astrophysical and cosmological scenarios. On the cosmological side, the largest anisotropy is expected to be generated through kinematic effects for which we analysed the sensitivity of the current datasets (NG15) and future experiments such as the SKA to the circular polarization kinematic dipole. Along the way, we have also highlighted the role played by the positions of the pulsars being observed, as well as the pulsar intrinsic noise properties, in the detection of the dipole and of the SGWB circular polarization. Although the sensitivity of current and near future datasets to the kinematic dipole is limited,  a close to maximal polarization may be a realistic target in the SKA era, or in related GW experiments based on astrometry. 

For the case of astrophysical SGWB anisotropy, the prospects
are much more optimistic. We have shown that the polarization anisotropy is a realistic target for near future experiments if the background is significantly polarised while a polarization level of a few percent may be detectable in the SKA era.

In our discussion, we made   optimistic hypothesis on the number and location of pulsars to be monitored, as well as their
noise properties. A more sophisticated
analysis, considering  a fully
 realistic treatment of the pulsar
noise curves, as
well as taking into account the role of cosmic variance \cite{Allen:2022dzg,Allen:2024bnk,Grimm:2024lfj,Agarwal:2024hlj} when
monitoring a finite number of pulsars, will be 
helpful towards  clarifying more accurately the
prospects 
of detecting circular polarization with PTA experiments.

\subsection*{Acknowledgments}
We are partially funded by the STFC grants ST/T000813/1 and ST/X000648/1. We also acknowledge the support of the Supercomputing Wales project, which is part-funded by the European Regional Development Fund (ERDF) via Welsh Government. For the purpose of open access, the authors have applied a Creative Commons Attribution licence to any Author Accepted Manuscript version arising.


{\small
\addcontentsline{toc}{section}{References}
\bibliographystyle{utphys}

\bibliography{refscirc}
}

\end{document}